\documentclass[sts]{imsart}

\RequirePackage{amsgen,amsmath,amstext,amsbsy,amsopn,amssymb,amscd,graphicx,cancel,needspace}

\usepackage{titlesec}
\titlespacing\section{0pt}{12pt plus 2pt minus 2pt}{5pt}
\titlespacing\subsection{0pt}{10pt plus 2pt minus 2pt}{3pt}

\startlocaldefs

\newcommand{\DS} {\displaystyle}
\newcommand{\TS} {\textstyle}

\newcommand{\as} {{\boldsymbol{P}}}
\newcommand{\const} {{\mathrm{const}}}
\renewcommand{\d} {{\mathrm{d}}}
\renewcommand{\l} {{{l}}}
\newcommand{\Reals} {{\mathrm{I}\!\mathrm{R}}}
\newcommand{\0} {{\boldsymbol{0}}}
\newcommand{\zero} {{\boldsymbol{0}}}
\newcommand{\one} {{\boldsymbol{1}}}
\newcommand{\x} {{\boldsymbol{x}}}
\newcommand{\y} {{\boldsymbol{y}}}
\renewcommand{\P} {{\boldsymbol{P}}}
\newcommand{\PYX} {{\boldsymbol{P}_{\!\!Y,\Xvec}}}
\newcommand{\PYgX} {{\boldsymbol{P}_{\!\!Y|\Xvec}}}
\newcommand{\PX} {{\boldsymbol{P}_{\!\!\Xvec}}}
\newcommand{\lean} {{{lean}}}
\newcommand{\lin}  {{{\!\!\:l\!\!\;i\!\!\;n}}}
\newcommand{\boot} {{{\!\!\:b\!\!\;o\!\!\;o\!\!\;t}}}
\newcommand{\sand} {{{\!\!\:s\!\!\:a\!\!\:n\!\!\:d}}}
\newcommand{\lmrob} {{{\!\!\:l\!\!\:m\!\!\:r\!\!\:o\!\!\:b}}}
\newcommand{\Phat} {{\boldsymbol{\hat{P}}}}

\newcommand{\E} {{\boldsymbol{E}}}
\newcommand{\Ehat} {\boldsymbol{\hat{E}}}
\newcommand{\V} {{\boldsymbol{V}}}
\newcommand{\Vhat} {\boldsymbol{\hat{V}}}
\newcommand{\Vlin} {{{\!\!\!lin}}}

\newcommand{\Vsand} {{{\!\!\!s\!\!\:a\!\!\:n\!\!\:d}}}

\newcommand{\AV} {{\boldsymbol{A\!V}}}
\newcommand{\AVhat}  {{{\boldsymbol{\hat{A\!V}}\!\!}}}
\newcommand{\AVboot}  {{{\boldsymbol{\hat{A\!V}}\!\!}_{boot}}}
\newcommand{\AVsand}  {{{\boldsymbol{\hat{A\!V}}\!\!\!}_{sand}}}

\newcommand{\AVlin} {{\boldsymbol{A\!V}_{\!\!\lin}}}

\newcommand{\AVlean} {{\boldsymbol{A\!V}_{\!\!\lean}}}

\newcommand{\RAV} {{\boldsymbol{R\!A\!V}}}

\newcommand{\RAVjhat} {{\boldsymbol{\hat{R\!A\!V}}_{\!\!j}}}
\newcommand{\D} {{\boldsymbol{D}}}
\newcommand{\Dhat} {{\hat{\boldsymbol{D}}}}
\newcommand{\I} {{\boldsymbol{I}}}
\renewcommand{\H} {{\boldsymbol{H}}}
\newcommand{\X} {{\bold{X}}}
\newcommand{\Xj} {{\X_{\!j}}}
\newcommand{\Xjm} {{\X_{\!j-1}}}
\newcommand{\Xjp} {{\X_{\!j+1}}}
\newcommand{\Xone} {{\X_{\!1}}}
\newcommand{\Xp} {{\X_{\!p}}}
\newcommand{\XX} {{{X}}}
\newcommand{\XXj} {{{X}_{\!j}}}
\newcommand{\XXjm} {{{X}_{\!j-1}}}
\newcommand{\XXjp} {{{X}_{\!j+1}}}
\newcommand{\XXone} {{{X}_{\!1}}}
\newcommand{\XXp} {{{X}_{\!p}}}
\newcommand{\Y} {{\bold{Y}}}

\renewcommand{\r} {{\boldsymbol{r}}}

\newcommand{\Dist} {{\cal D}}
\newcommand{\Norm} {{\cal N}}
\newcommand{\sig} {{\sigma}}
\newcommand{\eps} {{\epsilon}}

\newcommand{\Bnoise} {{\boldsymbol{\epsilon}}}
\newcommand{\rmse}{{m}}
\newcommand{\popres} {{\delta}}    

\newcommand{\Bpopres} {{\boldsymbol{\delta}}}
\newcommand{\nonlin} {{\eta}}
\newcommand{\Bnonlin} {{\boldsymbol{\eta}}}

\newcommand{\Bmu} {{\boldsymbol{\mu}}}
\newcommand{\Bbeta} {{\boldsymbol{\beta}}}

\newcommand{\tBbeta} {{\tilde{\boldsymbol{\beta}}}}
\newcommand{\Bbetahat} {{\boldsymbol{\hat{\beta}}}}
\newcommand{\BbetaX} {{\boldsymbol{\beta}(\X)}}
\newcommand{\BbetaP} {{\boldsymbol{\beta}(\P)}}
\newcommand{\Bbetaboot} {{{\boldsymbol{\beta}^*}}}
\newcommand{\Bbetadot} {{\Bbeta(\cdot)}}
\newcommand{\betaj} {{\beta_j}}
\newcommand{\betahat} {{\hat{\beta}}}
\newcommand{\betajhat} {{\hat{\beta}_{j}}}

\newcommand{\sigmahat} {{\hat{\sigma}}}
\newcommand{\Bsigma} {{\boldsymbol{\sigma}}}

\newcommand{\p} {{p}}
\newcommand{\pbar} {{\bar{p}}}

\newcommand{\argmin}{\mathrm{argmin}}

\newcommand{\SD} {{\boldsymbol{S\!D}}}
\newcommand{\SE} {{\boldsymbol{S\!E}}}
\newcommand{\SEhat} {{\boldsymbol{\hat{{S\!E}}}}}

\newcommand{\bull} {{\scriptscriptstyle \bullet}}
\newcommand{\hbull} {{\hat{\scriptscriptstyle \bullet}}}

\newcommand{\xvec} {{\vec{\boldsymbol{x}}}}

\newcommand{\Xvec} {{\vec{\boldsymbol{X}}}}
\newcommand{\Xveci} {{\vec{\boldsymbol{X}}_{\!i}}}
\newcommand{\Xvecone} {{\vec{\boldsymbol{X}}_{\!1}}}
\newcommand{\XvecN} {{\vec{\boldsymbol{X}}_{\!N}}}

\newcommand{\noj} {{{\!\scriptscriptstyle -}\!j}}

\newcommand{\Xsq}         {{X^2}}
\newcommand{\ft}          {{f_t}}
\newcommand{\fsq}         {{f^2}}
\newcommand{\fsqt}        {{f_t^2}}
\newcommand{\gt}          {{g_t}}

\newcommand{\gsqt}        {{g_t^2}}

\newcommand{\Xjadj}       {{X_{\!j \bull}}}
\newcommand{\Xjadjsq}     {{X_{\!j \bull}^{\,2}}}

\newcommand{\Xijadj}      {{X_{i,j \bull}}}

\newcommand{\Yadjj}       {{Y_{\! \bull \, \noj}}}

\newcommand{\XBjadj}       {{\X_{\!j \bull}}}

\newcommand{\Xjhadj}       {{X_{\!j \hbull}}}

\newcommand{\Xijhadj}      {{X_{i,j \hbull}}}

\newcommand{\XBjhadj}   {{\X_{\!j \hbull}}}

\newcommand{\YBhadjj}   {{\Y_{\!\! \hbull \, \noj}}}

\newcommand{\Bresid} {{\boldsymbol{r}}}

\newcommand{\siga} {{s_1}}
\newcommand{\sigb} {{s_2}}

\newcommand{\Btheta} {{\boldsymbol{\theta}}}

\newcommand{\Bpsi} {{\boldsymbol{\psi}}}
\newcommand{\Blambda} {{\boldsymbol{\lambda}}}

\newcommand{\A} {{\boldsymbol{A}}}
\renewcommand{\S} {{\boldsymbol{S}}}
\newcommand{\T} {{\boldsymbol{T}}}
\renewcommand{\a} {{\boldsymbol{a}}}

\renewcommand{\t} {{\boldsymbol{t}}}

\newcommand{\defn} {{\overset{\Delta}{=}}}

\newcommand{\xy}  {{x\textrm{-}y}}

\newcommand{\Tr}  {{'\,}}

\newcommand{\Pmax} {{\P\textrm{-}\!\max\,}}
\newcommand{\Pmin} {{\P\textrm{-}\!\min\,}}

\endlocaldefs

\begin{document}


\begin{frontmatter}

  \title{ Models as Approximations I: Consequences Illustrated with Linear Regression}
  \runtitle{Models as Approximations}

\begin{aug}

  \author{\fnms{Andreas}
          \snm{Buja}\thanksref{t1,m1}\ead[label=e1]{buja.at.wharton@gmail.com}},
  \author{\fnms{Richard}
          \snm{Berk}\thanksref{m1}\ead[label=e2]{berkr@wharton.upenn.edu}},
  \author{\fnms{Lawrence}
          \snm{Brown}\thanksref{t1,m1}\ead[label=e3]{lbrown@wharton.upenn.edu}},
  \author{\fnms{Edward}
          \snm{George}\thanksref{t3,m1}\ead[label=e4]{edgeorge@wharton.upenn.edu}},
  \author{\fnms{Emil}
          \snm{Pitkin}\thanksref{t1,m1}\ead[label=e5]{pitkin@wharton.upenn.edu}},
  \author{\fnms{Mikhail}
          \snm{Traskin}\thanksref{m2}
          },
  \author{\fnms{Linda}
          \snm{Zhao}\thanksref{t1,m1}\ead[label=e6]{lzhao@wharton.upenn.edu}}
  \and
  \author{\fnms{Kai}
          \snm{Zhang}\thanksref{t1,m3}\ead[label=e7]{zhangk@email.unc.edu}}

  \thankstext{t1}{Supported in part by NSF Grant DMS-10-07657 and DMS-1310795.}  
  \thankstext{t3}{Supported in part by NSF Grant DMS-14-06563.}  

  \runauthor{A. Buja et al.}

  \affiliation{Wharton -- University of Pennsylvania\thanksmark{m1}
               and
               Citadel\thanksmark{m2}
               and
               UNC at Chapel Hill\thanksmark{m3}}

  \address{Statistics Department,
           The Wharton School, University of Pennsylvania,
           400 Jon M. Huntsman Hall, 3730 Walnut Street,
           Philadelphia, PA 19104-6340
           \printead{e1}.}
  \address{-- Citadel, Chicago.}
  \address{-- Dept. of Statistics \& Operations Research,
           306 Hanes Hall, CB\#3260,
           The University of North Carolina at Chapel Hill,
           Chapel Hill, NC 27599-3260.}

\end{aug}

\maketitle


\begin{abstract}

  In the early 1980s Halbert White inaugurated a
  ``model-robust'' form of statistical inference based on the
  ``sandwich estimator'' of standard error.  This estimator is known
  to be ``heteroskedasticity-consistent'', but it is less well-known
  to be ``nonlinearity-consistent'' as well.  Nonlinearity, however,
  raises fundamental issues because in its presence regressors are not
  ancillary, hence can't be treated as fixed.
  The consequences are deep: (1)~population slopes need to be
  re-interpreted as statistical functionals obtained from OLS fits to
  largely arbitrary joint $\xy$~distributions; (2)~the meaning of
  slope parameters needs to be rethought; (3)~the regressor
  distribution affects the slope parameters; (4)~randomness of the
  regressors becomes a source of sampling variability in slope
  estimates; (5)~inference needs to be based on model-robust standard
  errors, including sandwich estimators or the $\xy$~bootstrap.  In
  theory, model-robust and model-trusting standard errors can deviate
  by arbitrary magnitudes either way.  In practice, significant
  deviations between them can be detected with a diagnostic test.
\end{abstract}

\begin{keyword}[class=AMS]
\kwd[Primary ]{62J05}      
\kwd{62J20}                
\kwd{62F40}                
\kwd[; secondary ]{62F35}  
\kwd{62A10}                
\end{keyword}

\begin{keyword}
\kwd{Ancillarity of regressors}
\kwd{Misspecification}
\kwd{Econometrics}
\kwd{Sandwich estimator}
\kwd{Bootstrap}
\end{keyword}

\end{frontmatter}


\section{Introduction}
\label{sec:intro}

Halbert White's basic sandwich estimator of standard error for OLS can
be described as follows: In a linear model with regressor matrix
$\X_{\!N\!\times\!(p\!+\!1)}$ and response vector $\y_{N\!\times\!1}$,
start with the familiar derivation of the covariance matrix of the OLS
coefficient estimate $\Bbetahat$, but allow heteroskedasticity,
$\V[\y|\X] \!=\! \D$ diagonal:
\begin{equation} \label{eq:sandwich-true}
  \V[\,\Bbetahat\,|\X] = \V[ (\X\Tr\!\X)^{-1} \X\Tr \y \,|\X]
                       = (\X\Tr\!\X)^{-1} ( \X\Tr \D \X ) (\X\Tr\!\X)^{-1}.
\end{equation}
The right hand side has the characteristic ``sandwich'' form,
$(\X\Tr\!\X)^{-1}$ forming the ``bread'' and $\X\Tr \D \X$ the
``meat.''  Although this sandwich formula does not look actionable for
standard error estimation because the variances
$\D_{ii} \!=\! \sig_i^2$ are not known, White showed that
\eqref{eq:sandwich-true} can be estimated asymptotically correctly.
If one estimates $\sig_i^2$ by squared residuals $r_i^2$, each $r_i^2$
is not a good estimate, but the averaging implicit in the ``meat''
provides an asymptotically valid estimate:\footnote{
  This sandwich estimator
  is only the simplest version of its kind.  Other versions were
  examined, for example, by MacKinnon and White
  (1985)\nocite{ref:MW-1985} and Long and Ervin
  (2000)\nocite{ref:LE-2000}.  Some forms are pervasive in Generalized
  Estimating Equations (GEE; Liang and Zeger 1986; Diggle et
  al. 2002)\nocite{ref:LZ-1986}\nocite{ref:DHLZ-2002} and in the
  Generalized Method of Moments (GMM; Hansen 1982; Hall
  2005)\nocite{ref:Hansen-1982}\nocite{ref:Hall-2005}.  }
\begin{equation} \label{eq:sandwich-est}
  \Vhat_\Vsand[\,\Bbetahat\,] ~~\defn~~ (\X\Tr\!\X)^{-1} ( \X\Tr \Dhat \X ) (\X\Tr\!\X)^{-1} ,
\end{equation}
where $\Dhat$ is diagonal with $\Dhat_{ii} \!=\! r_i^2$.  Standard
error estimates are obtained by
$\SEhat_{\sand}[\,\betajhat\,] \!=\! \Vhat_\Vsand[\,\Bbetahat\,]_{jj}^{1/2}$.
They are asymptotically valid even if the responses are
heteroskedastic, hence the term ``Heteroskedasticity-Consistent
Covariance Matrix Estimator'' in the title of one of White's
(1980b\nocite{ref:White-1980b}) famous articles.

Lesser known is the following deeper result in one of White's
(1980a\nocite{ref:White-1980a}, p.~162-3) less widely read articles:
the sandwich estimator of standard error is asymptotically correct
even in the presence of nonlinearity:\footnote{ The term
  ``nonlinearity'' is meant in the sense of first order model
  misspecification.  A different meaning of ``nonlinearity'', {\em
    not} intended here, occurs when the regressor matrix $\X$ contains
  multiple columns that are functions (products, polynomials,
  B-splines,~...) of underlying independent variables.  One needs to
  distinguish between ``regressors'' and ``independent variables'':
  Multiple regressors may be functions of one or more independent
  variable(s).  }
\begin{equation} \label{eq:misspecification-1st-order}
  \E[\,\y\,|\X] ~\neq~ \X \Bbeta~~~~{\rm for~all}~\Bbeta.
\end{equation}
The term ``heteroskedasticity-consistent'' is an unfortunate choice as
it obscures the fact that the same estimator of standard error is also
``nonlinearity-consistent'' {\em when the regressors are treated as
  random}.  The sandwich estimator of standard error is therefore
``model-robust'' not only against second order model violations but
first order violations as well.  Because of the relative obscurity of
this important fact we will pay considerable attention to its
implications.  In particular we will show how {\em nonlinearity}
``conspires'' with {\em randomness of the regressors}
\begin{itemize}
\item[(1)] to make slopes dependent on the regressor distribution and
\item[(2)] to generate sampling variability, even in the absence of noise in the response.
\end{itemize}
For an intuitive grasp of these effects, the reader may peruse
Figure~\ref{fig:nonlinearity-population} for effect~(1) and
Figure~\ref{fig:no-error} for effect~(2).\footnote{A more striking
  illustration of effect (2) in the form of an animation is available
  to users of the {\em {\bf R} Language}~(2008)\nocite{ref:R-2008} by
  executing the following line of code:
  \\
  \centerline{\tt\footnotesize
    source("http://stat.wharton.upenn.edu/\~{}buja/src-conspiracy-animation2.R")
  } }


From the sandwich estimator \eqref{eq:sandwich-est}, the usual {\em
  model-trusting} estimator is obtained by collapsing the sandwich
form using homoskedasticity, $\Dhat = \sigmahat \I$:
\begin{equation*}
  \Vhat_\Vlin[\,\Bbetahat\,] ~~\defn~~ (\X\Tr\!\X)^{-1} \sigmahat^2 ,~~~~ \sigmahat^2 = \|\r\|^2/(N\!\!-\!p\!-\!1) .
\end{equation*}
This yields finite-sample unbiased squared standard error estimators
$\SEhat_{\lin}^2[\,\betajhat\,] \!=\! \Vhat_\Vlin[\,\Bbetahat\,]_{jj}$
if the model is first and second order correct:
$\E[\y\,|\X] \!=\! \X \Bbeta$ (linearity) and
$\V[\y\,|\X] \!=\! \sigma^2 \I_N$ (homoskedasticity).  Assuming
distributional correctness (Gaussian errors), one obtains
finite-sample correct tests and confidence intervals.

The corresponding tests and confidence intervals based on the sandwich
estimator have only an asymptotic justification, but their asymptotic
validity holds under much weaker assumptions.  In fact, it may rely on
no more than the assumption that the rows $(y_i,\xvec_i')$ of the data
matrix $(\y,\X)$ are iid~samples from a joint multivariate
distribution subject to some technical conditions. Thus sandwich-based
theory provides asymptotically correct inference that is {\em
  model-robust}.  The question then arises what model-robust inference
is about: When no model is assumed, {\em what are the parameters}, and
{\em what is their meaning}?

Discussing these questions is a first goal of this article.  An
established answer is that parameters can be re-interpreted as {\em
  statistical functionals} $\BbetaP$ defined on a large nonparametric
class of joint distributions $\P=\P(dy,d\xvec)$ through best
approximation (Section~\ref{sec:population-framework}), sometimes
called ``projection.''  The sandwich estimator produces then
asymptotically correct standard errors for the slope functionals
$\betaj(\P)$ (Section~\ref{sec:estimation}).  Vexing is the
question of the meaning of slopes in the presence of nonlinearity as
the standard interpretations no longer apply.  We will propose
interpretations that draw on the notions of {\em case-wise} and {\em
  pairwise slopes} after linear adjustment
(Section~\ref{sec:ave-slopes}).

A second goal of this article is to discuss why the regressors should
be treated as random.  Based on an ancillarity argument,
model-trusting theories tend to condition on the regressors and hence
treat them as fixed (Cox and Hinkley 1974, p.~32f\nocite{ref:CH-1974},
Lehmann and Romano 2008, p.~395ff\nocite{ref:LR-2008}).  However, it
will be shown that under misspecification ancillarity of the
regressors is violated (Section~\ref{sec:non-ancillarity}).  Here are
some implications:
\begin{itemize} \itemsep 0em
\item Population parameters $\Bbeta(\P)$, now interpreted as
  statistical functionals, depend on the distribution of the
  regressors.  Thus it matters where the regressors fall.  The reason
  is intuitive: When models are approximations, it matters where the
  approximation is made; see
  Figure~\ref{fig:nonlinearity-population}.
\item A natural intuition fails, caused by misleading terminology:
  Nonlinearity --- sometimes called ``model bias'' --- does {\em not}
  primarily cause bias in estimates $\Bbeta(\Phat)$.  It causes
  sampling variability of order~$N^{-1/2}$, thereby rivaling
  error/noise as a source of sampling variability
  (Section~\ref{sec:randomX-plus-nonlinearity}).
\item A second intuition fails: While it is correct that an inference
  guarantee conditional on the regressors implies a marginal inference
  guarantee, this principle is inapplicable because the premise is
  false --- {\em under misspecification there is no inference
    guarantee conditional on the regressors.}  The reason is that
  inference theories that treat regressors as fixed are incapable of
  correctly accounting for misspecification.
\end{itemize}
All three implications hold in great generality, but in this article
they will be worked out for OLS linear regression to achieve the
greatest degree of lucidity.


A third goal of this article is to argue in favor of the ``$\xy$
bootstrap'' which resamples observations $(\xvec_i',y_i)$.  The better
known ``residual bootstrap'' resamples residuals~$r_i$ and thereby
assumes a linear response surface and exchangeable errors.  There
exists theory to justify both (Freedman
(1981)\nocite{ref:Freedman-1981} and Mammen
(1993)\nocite{ref:Mammen-1993}, for example), but only the $\xy$
bootstrap is model-robust and solves the same problem as the sandwich
estimator.\footnote{Note David Freedman's
  (1981)\nocite{ref:Freedman-1981} surprise when he inadvertently
  discovered the same assumption-lean validity of the $\xy$ bootstrap
  (ibid.~top of p.~1220).}  In Part~II (Buja et
al.~2018\nocite{ref:Buja-et-al-2018b}), it will be shown that the
sandwich estimator is a limiting case of the $\xy$ bootstrap.

A fourth goal of this article is to practically
(Section~\ref{sec:data-example}) and theoretically
(Section~\ref{sec:AVs}) compare model-robust and model-trusting
estimators of standard error in the case of OLS linear regression.  To
this end we define a ratio of asymptotic variances --- ``$\RAV$'' for
short --- that describes the discrepancies between the two standard
errors in the asymptotic limit.

A fifth goal is to estimate the $\RAV$ for use as a test statistic.
We derive an asymptotic null distribution to test for model deviations
that invalidate the usual standard error of a specific coefficient.
The resulting ``misspecification test'' differs from other such tests
in that it answers the question of discrepancies among standard errors
directly and separately for each coefficient
(Section~\ref{sec:sandwich-adjusted-and-RAV-test}).

A final goal is to briefly discuss issues with sandwich estimators
(Section~\ref{sec:robustness}): They can be inefficient when models
are correctly specified.  We additionally point out that they are
non-robust to heavy tails in the joint $\xy$ distribution.  To make
sense of this observation, the following distinctions are needed:
(1)~classical robustness to heavy tails is distinct from model
robustness to first and second order model misspecifications; (2)~at
issue is not robustness (in either sense) of parameter estimates but
of standard errors.  It is the latter we examine here.


Throughout we use precise notation for clarity, yet this article is
not very technical.  Many results are elementary, not new, and stated
without regularity conditions.  Readers may browse the tables and
figures and read associated sections that seem most germane.
Important notations are shown in boxes.

The present article is limited to OLS linear regression, both for
populations and for data.  The case permits explicit calculations and
lucid interpretations.  Part II (Buja et
al. 2018\nocite{ref:Buja-et-al-2018b}) will treat arbitrary
regressions at the cost of reduced lucidity.  It will also propose new
types of diagnostics.


The idea that models are approximations and hence generally
``misspecified'' to a degree has a long history, most famously
expressed by Box (1979)\nocite{ref:Box-1979}.  We prefer to quote Cox
(1995)\nocite{ref:Cox-1995}: ``it does not seem helpful just to say
that all models are wrong.  The very word model implies simplification
and idealization.''  The history of inference under misspecification
can be traced to Cox (1961, 1962), Eicker
(1963)\nocite{ref:Eicker-1963}, Berk(1966,
1970)\nocite{ref:Berk-1966,ref:Berk-1970}, Huber
(1967)\nocite{ref:Huber-1967}, before being systematically elaborated
by White's articles (1980a, 1980b, 1981, 1982, among others),
\nocite{ref:White-1980a,ref:White-1980b,ref:White-1981,ref:White-1982}
capped by a monograph (White 1994)\nocite{ref:White-1994}.  A
wide-ranging discussion by Wasserman (2011)\nocite{ref:Wasserman-2011}
calls for ``Low Assumptions, High Dimensions.''  A book by Davies
(2014)\nocite{ref:Davies-2014} elaborates the idea of adequate models
for a given sample size.  We, the present authors, got involved with
this topic through our work on post-selection inference (Berk et
al. 2013)\nocite{ref:Berk-et-al-2013} because the results of model
selection should certainly not be assumed to be ``correct.''  We
compared the obviously model-robust standard errors of the $\xy$
bootstrap with the usual ones of linear models theory and found the
discrepancies illustrated in Section~\ref{sec:data-example}.
Attempting to account for these discrepancies became the starting
point of the present article.


\section{Discrepancies between Standard Errors Illustrated}
\label{sec:data-example}

Table~\ref{tab:LA} shows regression results for a dataset consisting
of a sample of 505 census tracts in Los Angeles that has been used to
relate the local number of homeless ($Y$) to covariates for
demographics and building usage (Berk et
al. 2008)\nocite{ref:BKY-2008}.\footnote{ The response is the raw
  number of homeless in a census tract.  The tracts do not differ by
  magnitudes and, according to experts, size effects seem minor.  The
  homeless tend to clump in certain areas within census tracts, and it
  is thought that the regressors describe features of the tracts that
  make them magnets for the homeless.  Finally, policy makers are
  accustomed to thinking in counts, not percentages.}  We do not
intend a careful modeling exercise but show the raw results of linear
regression to illustrate the degree to which discrepancies can arise
among three types of standard errors: $\SE_\lin$ from linear models
theory, $\SE_\boot$ from the $\xy$ bootstrap ($N_\boot = 100,000$) and
$\SE_\sand$ from the sandwich estimator (according to MacKinnon and
White's (1985)\nocite{ref:MW-1985} HC2 proposal).  Ratios of standard
errors that are far from +1 are shown in bold font.

\begin{table}
\resizebox{\textwidth}{!}{\tt
\begin{tabular}{l|rrrrrrrrrr}
                  &$\betajhat$ & $\SE_\lin$ & $\SE_\boot$ & $\SE_\sand$ & $\frac{\SE_\boot}{\SE_\lin}$ & $\frac{\SE_\sand}{\SE_\lin}$ & $\frac{\SE_\sand}{\SE_\boot}$ & $t_\lin$ & $t_\boot$ & $t_\sand$ \\
\hline
\texttt{Intercept}         &  0.760 & 22.767 & 16.505 & 16.209 & \bf 0.726 & \bf 0.712 & 0.981 &  0.033 &  0.046 &  0.047 \\
\texttt{MedianIncome (\$K)}& -0.183 &  0.187 &  0.114 &  0.108 & \bf 0.610 & \bf 0.576 & 0.944 & -0.977 & -1.601 & -1.696 \\
\texttt{PercVacant}        &  4.629 &  0.901 &  1.385 &  1.363 & \bf 1.531 & \bf 1.513 & 0.988 &  5.140 &  3.341 &  3.396 \\
\texttt{PercMinority}      &  0.123 &  0.176 &  0.165 &  0.164 &     0.937 &     0.932 & 0.995 &  0.701 &  0.748 &  0.752 \\
\texttt{PercResidential}   & -0.050 &  0.171 &  0.112 &  0.111 & \bf 0.653 & \bf 0.646 & 0.988 & -0.292 & -0.446 & -0.453 \\
\texttt{PercCommercial}    &  0.737 &  0.273 &  0.390 &  0.397 & \bf 1.438 & \bf 1.454 & 1.011 &  2.700 &  1.892 &  1.857 \\
\texttt{PercIndustrial}    &  0.905 &  0.321 &  0.577 &  0.592 & \bf 1.801 & \bf 1.843 & 1.023 &  2.818 &  1.570 &  1.529
\end{tabular}
}
\caption{\it LA Homeless Data: Comparison of Standard Errors.  See
  also Table~\ref{tab:Boston} in the Appendix for the Boston Housing Data.}
\label{tab:LA}
\end{table}

The ratios $\SE_\sand/\SE_\boot$ show that the sandwich and bootstrap
estimators are in good agreement.  Not so for the linear models
estimates: we have $\SE_\boot, \SE_\sand > \SE_\lin$ for the
regressors \texttt{PercVacant}, \texttt{PercCommercial} and
\texttt{PercIndustrial}, and $\SE_\boot, \SE_\sand < \SE_\lin$ for
\texttt{Intercept}, \texttt{MedianIncome (\$K)},
\texttt{PercResidential}.  Only for \texttt{PercMinority} is
$\SE_\lin$ off by less than 10\% from $\SE_\boot$ and $\SE_\sand$.
The discrepancies affect outcomes of some of the $t$-tests: Under
linear models theory the regressors \texttt{PercCommercial} and
\texttt{PercIndustrial} have sizable $t$-values of $2.700$ and
$2.818$, respectively, which are reduced to unconvincing values below
$1.9$ and $1.6$, respectively, if the $\xy$ bootstrap or the sandwich
estimator are used.  On the other hand, for \texttt{MedianIncome (\$K)}
the $t$-value $-0.977$ from linear models theory becomes borderline
significant with the bootstrap or sandwich estimator if the plausible
one-sided alternative with negative sign is used.

A similar exercise with fewer discrepancies but similar
conclusions is shown in Appendix~\ref{app:Boston} for the Boston
Housing data.

{\bf Conclusions: }
(1)~$\SE_\boot$ and $\SE_\sand$ are in substantial agreement;
(2)~$\SE_\lin$ on the one hand and $\{\SE_\boot,\SE_\sand\}$ on the
other hand can have substantial discrepancies; (3)~the discrepancies
are specific to regressors.


\section{The Population Framework for Linear OLS}
\label{sec:population-framework}

As noted earlier, model-robust inference needs a target of estimation
that is well-defined outside the linear working model.  To this end we
need notation for data distributions that are free of model
assumptions, essentially relying on iid~sampling of $\xy$~tuples.
Subsequently OLS parameters can be introduced as statistical
functionals of these distributions through best linear approximation.
This is sometimes called ``projection'', meaning that the
assumption-free data distribution is ``projected'' to the ``nearest''
distribution in the working model.


\subsection{Populations for OLS Linear Regression}
\label{sec:populations}

In an assumption-lean, model-robust population framework for OLS
linear regression with random regressors, the ingredients are {\em
  regressor random variables} $X_1$, ..., $X_p$ and a {\em response
  random variable}~$Y$.  For now the only assumption is that they are
all numeric and have a joint distribution, written as
\begin{equation*}
  \P ~=~ \P(\d y, \d x_1,...,\d x_p) .
\end{equation*}
Data will consist of iid~multivariate samples from this joint
distribution (Section~\ref{sec:estimation}).  {\bf\em No working
  model for $\P$ will be assumed}.

It is convenient to add a fixed regressor 1 to accommodate an
intercept parameter; we may hence write
\begin{equation*} 
  \Xvec ~=~ (1,X_1,...,X_p)\Tr
\end{equation*}
for the {\em column random vector} of the regressor variables, and
$\xvec = (1,x_1,...,x_p)\Tr$ for its values.  We further write
\[
\PYX = \P,~~~~ \PYgX,~~~~ \PX,
\]
for, respectively, the joint distribution of $(Y,\Xvec)$, the
conditional distribution of $Y$ given $\Xvec$, and the marginal
distribution of $\Xvec$.  These denote {\em actual} data distributions,
free of assumptions of a working model.

All variables will be assumed to be square integrable.  Required is
also that $\E[ \Xvec \Xvec\Tr ]$ is full-rank, but permitted are
nonlinear degeneracies among regressors as when they are functions of
underlying independent variables such as in polynomial or B-spline
regression or product interactions.

\begin{figure}[!t]
  \centering
  \includegraphics[height=3.5in]{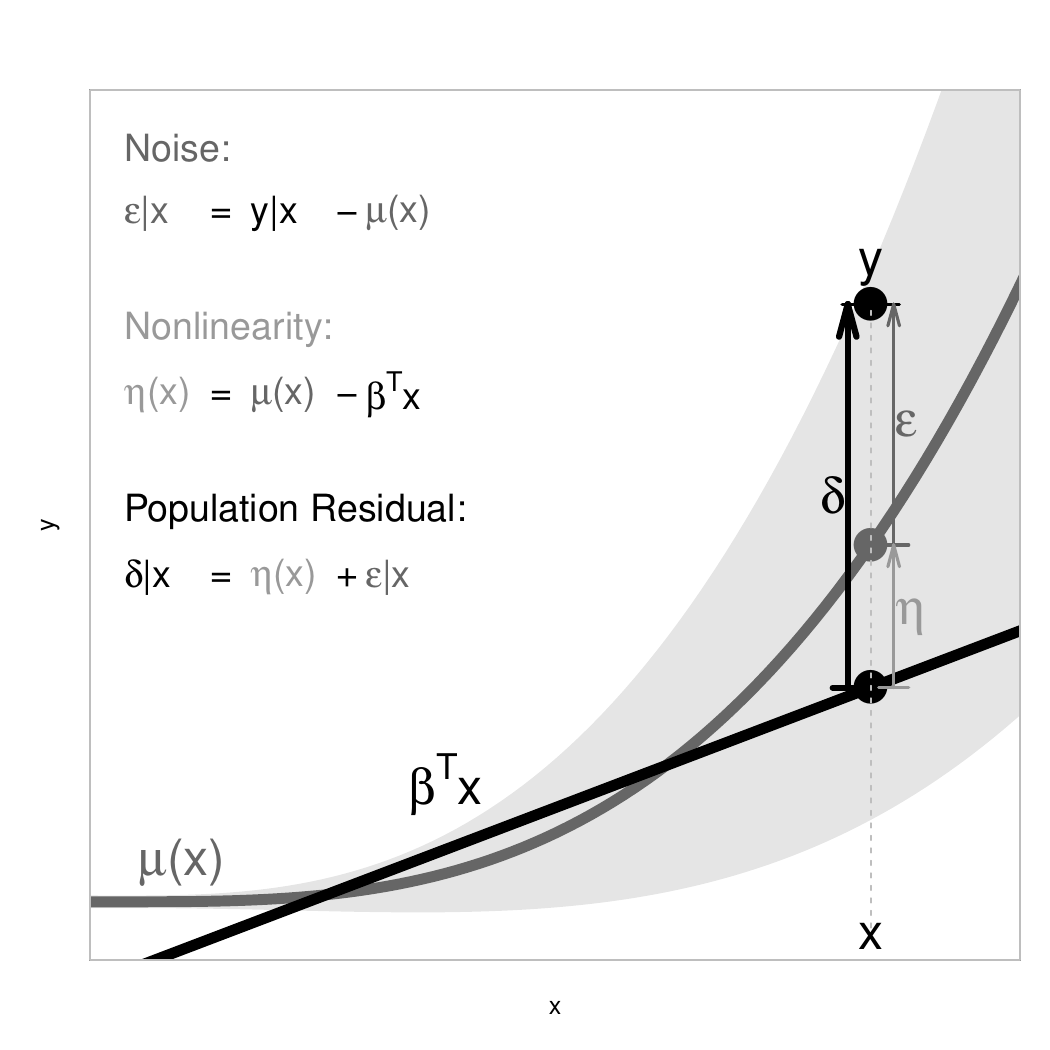}
  \vspace{-.1in}
  \caption{\it Illustration of the decomposition \eqref{eq:decomp} for linear OLS.}
  \label{fig:decomp}
\end{figure}


\subsection{Targets of Estimation: The OLS Statistical Functional}
\label{sec:targets-of-estimation}

We write any function $f(X_1,...,X_p)$ of the regressors as
$f(\Xvec)$.  We will need notation for the ``true response surface''
$\mu(\Xvec)$, which is the conditional expectation of $Y$ given
$\Xvec$ and the best $L_2(\P)$ approximation to $Y$ among functions
of~$\Xvec$.  It is {\bf\em not} assumed to be linear in~$\Xvec$:
\begin{equation*}
  \mu(\Xvec) ~\defn~
  \E[\,Y\,|\,\Xvec\,]
  ~=~
  \argmin_{f(\Xvec) \in L_2(\P)} \, \E[ (Y - f(\Xvec))^2 ] \,.
\end{equation*}
The main definition concerns {\bf\em the best population linear
  approximation} to $Y$, which is the linear function
$\l(\Xvec) = \Bbeta\Tr \Xvec$ with coefficients $\Bbeta = \BbetaP$
given by
\begin{equation*}
  \arraycolsep=2pt
  \def\arraystretch{2.0}
  \begin{array}{|lclcl|}
    \hline
    ~~~~\BbetaP~ &\defn& \argmin_{\Bbeta \in \Reals^{p+1}} \, \E[ (Y\!-\!\Bbeta\Tr \Xvec )^2 ] ~~~~~
                 &=&  \E[\Xvec \Xvec\Tr]^{-1} \E[\Xvec Y\,]
    \\
                 &=&  \argmin_{\Bbeta \in \Reals^{p+1}} \, \E[ (\mu(\Xvec)\!-\!\Bbeta\Tr \Xvec )^2 ]
                 &=&  \E[\Xvec \Xvec\Tr]^{-1} \E[\Xvec \mu(\Xvec)\,] . ~~~~
    \\
    \hline
  \end{array}
\end{equation*}
Both right hand expressions follow from the population normal
equations:
\begin{equation} \label{eq:normal-equations}
\E[\Xvec \Xvec\Tr] \, \Bbeta - \E[\Xvec Y]  ~=~
\E[\Xvec \Xvec\Tr] \, \Bbeta - \E[\Xvec \mu(\Xvec)] ~=~ \0 .
\end{equation}

The population coefficients
$\BbetaP = (\beta_0(\P),\beta_1(\P),...,\beta_p(\P))\Tr$ form a
{\bf\em vector statistical functional}, $\P \mapsto \BbetaP$, defined
for a large class of joint data distributions
$\P\!=\!\PYX$.  If the response surface under $\P$ happens
to be linear, $\mu(\Xvec) \!=\! \tilde{\Bbeta}\Tr \Xvec$, as it is for
example under a Gaussian linear model,
$Y|\Xvec \sim \Norm(\tilde{\Bbeta}\Tr \Xvec, \sigma^2)$, then
$\BbetaP \!=\! \tilde{\Bbeta}$.  The statistical functional is
therefore a natural extension of the traditional meaning of a model
parameter, justifying the notation $\Bbeta \!=\! \BbetaP$.  The point
is, however, that $\Bbetadot$ is defined even when linearity does not
hold.  (Depending on the context, we may write $\Bbeta$ to
mean~$\Bbeta(\P)$.)


\subsection{The Noise-Nonlinearity Decomposition for Population OLS}
\label{sec:noise-nonlinearity-decomposition}

The response $Y$ has the following canonical decompositions:
\begin{equation} \label{eq:decomp}
  \begin{array}{ccccc}
    Y &=& \Bbeta\Tr \Xvec &+& \underbrace{(\mu(\Xvec) - \Bbeta\Tr \Xvec)} ~+~ \underbrace{(Y - \mu(\Xvec))} \\[1em]
      &=& \Bbeta\Tr \Xvec &+&            \underbrace{\nonlin(\Xvec)  ~~~~~~+~~~~~~~~     \eps}        \\[1em]
      &=& \Bbeta\Tr \Xvec &+&                             \popres        \\
  \end{array}
\end{equation}
We call $\eps = \eps|\Xvec$ the {\em noise} and
$\nonlin = \nonlin(\Xvec)$ the {\em nonlinearity}$\,$\footnote{The term
  ``nonlinearity'' has two meanings, related to each other.  ``The/a
  nonlinearity'' refers to $\nonlin(\xvec)$, but ``presence of
  nonlinearity'' is a property of~$\mu(\xvec)$.}, while for $\popres$
there is no standard term, so ``{\em population residual}'' may
suffice; see Table~\ref{eq:decomp-popres} and Figure~\ref{fig:decomp}.
Important to note is that \eqref{eq:decomp} is a {\em decomposition,
  not a model assumption}.  In a model-robust framework there is no
notion of ``error term'' in the usual sense; its place is taken by the
population residual $\popres$ which satisfies few of the usual
assumptions made in generative models.  It naturally decomposes into a
systematic component, the nonlinearity $\nonlin = \nonlin(\Xvec)$, and
a random component, the noise $\eps = \eps|\Xvec$.  Model-trusting
linear modeling, before conditioning on $\Xvec$, must assume
$\nonlin(\Xvec) \overset{\P}{=} 0$ and $\eps$ to have the same
$\Xvec$-conditional distribution in all of regressor space, that is,
to be independent of $\Xvec$.  No such assumptions are made here.
What is left are orthogonalities satisfied by $\nonlin$ and $\eps$ in
relation to~$\Xvec$.  If we call independence ``strong-sense
orthogonality'', we have instead
\begin{equation} \label{eq:orthogonalities}
  \begin{array}{ll}
    \textrm{weak-sense orthogonality:} ~~~~~ \nonlin \perp \Xvec        &
    \!\!( \E[ \nonlin \!\cdot\!\! X_j ] = 0   ~~ \forall j \!=\! 0,1,...,p ),
    \\[0.3em]
    \textrm{medium-sense orthogonality:}~    \eps \perp L_2(\PX) &
    \!\!( \E[\eps \!\cdot\!\! f(\!\Xvec\!)] = 0 ~ \forall f \!\in\! L_2(\PX) ).
  \end{array}
\end{equation}
These are not assumptions but consequences of population OLS and the
definitions.  Because of the inclusion of an intercept ($j\!=\!0$ and
$f\!=\!1$, respectively), both the nonlinearity and noise are
marginally centered: $\E[\,\nonlin\,] = \E[\,\eps\,] = 0$.  Importantly, it
also follows that $\eps \perp \nonlin(\Xvec)$ because $\nonlin$ is
just some $f \!\in\! L_2(\PX)$.

\smallskip

In what follows we will need the following natural definitions:
\begin{itemize} \itemsep 0.5em
\item {\bf\em Conditional noise variance}: The noise $\eps$, not
  assumed homoskedastic, can have arbitrary conditional distributions
  $\P(d \eps | \Xvec\!=\!\xvec)$ for different $\xvec$ except for
  conditional centering and finite conditional variances.  Define:
  \begin{equation} \label{eq:sigma2}
    \sig^2(\Xvec) ~~\defn~~ \V[\,\eps \,| \Xvec] ~=~ \E[\, \eps^2 \,|\, \Xvec] ~\overset{\as}{<}~ \infty.
  \end{equation}
  When we use the abbreviation $\sig^2$ we will mean
  $\sig^2 = \sig^2(\Xvec)$ as we will never assume homoskedasticity.
\item {\bf\em Conditional mean squared error}: This is the conditional
  MSE of $Y$ w.r.t.~the population linear approximation
  $\Bbeta\Tr \Xvec$.  Its definition and bias-variance decomposition
  are:
  \begin{equation} \label{eq:mse}
    \rmse^2(\Xvec) ~~\defn~~ \E[\, \popres^2 \,|\, \Xvec]
                   ~=~ \nonlin^2(\Xvec) + \sig^2(\Xvec) .
  \end{equation}
  The right hand side follows from
  $\popres = \nonlin\!+\!\eps$ and $\eps \perp \nonlin(\Xvec)$ noted
  after~\eqref{eq:orthogonalities}.
\end{itemize}
In the above definitions and statements, randomness of the regressor
vector $\Xvec$ has started to play a role.  The next section will
discuss a crucial role of the marginal regressor distribution~$\PX$.



\begin{table}
\begin{equation*}
  \arraycolsep=1.5pt
  \def\arraystretch{1.7}
  \begin{array}{|clll|}
    \hline
    ~~~
    \nonlin &=~ \mu(\Xvec) - \Bbeta\Tr \Xvec &~=~ \nonlin(\Xvec),
    & \textrm{~~~~~~{\bf\em nonlinearity}},
    \\
    ~~~
    \eps    &=~ Y \!\!- \mu(\Xvec), &
    &  \textrm{~~~~~~{\bf\em noise}},
    \\
    ~~~
    \popres &=~ Y\!\!-\!\Bbeta\Tr \Xvec &~=~ \nonlin \,+\, \eps,
    & \textrm{~~~~~~{\bf\em population residual}},~
    \\
    ~~~
    \mu(\Xvec) &=~ \Bbeta\Tr \Xvec + \nonlin(\Xvec)&~~~
    & \textrm{~~~~~~{\bf\em response surface}},~
    \\
    ~~~
    Y &=~ \Bbeta\Tr \Xvec + \nonlin(\Xvec) + \eps &~=~ \Bbeta\Tr \Xvec + \popres
    & \textrm{~~~~~~{\bf\em response}}.~
    \\
    \hline
  \end{array}
\end{equation*}
\caption{Random variables and their canonical decompositions.}
\label{eq:decomp-popres}
\end{table}


\smallskip

\section{Broken Regressor Ancillarity I:  ~~~~~~~~~~~~~~~~~~~~~ Nonlinearity and Random $\X$ Jointly Affect Slopes}
\label{sec:non-ancillarity}



\smallskip

\subsection{Misspecification Destroys Regressor Ancillarity}
\label{sec:ancillarity-breakdown}

Conditioning on the regressors and treating them as fixed when they
are random has historically been justified with the ancillarity
principle.  Regressor ancillarity is a property of working models
$p(y\,|\,\xvec;\,\Btheta)$ for the conditional distribution of
$Y|\Xvec$, where $\Btheta$ is the parameter of interest in the usual
meaning of a parametric model.  Because we treat $\Xvec$ as random,
the assumed joint distribution of $(Y,\Xvec)$~is
\begin{equation*}
  p(y,\xvec;\,\Btheta) ~=~ p(y\,|\,\xvec;\,\Btheta) \; p(\xvec) ,
\end{equation*}
where $p(\xvec)$ is the unknown marginal regressor distribution,
acting as a ``non-parametric nuisance parameter.''  Ancillarity of
$p(\xvec)$ in relation to $\Btheta$ is immediately recognized by
forming likelihood ratios,
\begin{equation*}
p(y,\xvec;\,\Btheta_1) / p(y,\xvec;\,\Btheta_2) ~=~
p(y\,|\,\xvec;\,\Btheta_1) / p(y\,|\,\xvec;\,\Btheta_2) ,
\end{equation*}
which are free of $p(\xvec)$, detaching the regressor distribution
from inference about the parameter $\Btheta$.  (For more on
ancillarity, see Appendix~\ref{app:ancillarity}.)  This logic is valid
if $p(y\,|\,\xvec;\,\Btheta)$ correctly describes the actual
conditional regressor distribution $\PYgX$ for some $\Btheta$.  If
this is not the case, the ancillarity argument does not apply.

To pursue the consequences of non-ancillarity, one needs to consider
$\PYgX$ not in the working model and interpret parameters as
statistical functionals:

\smallskip

\vspace{.05in}\noindent{\bf Proposition \ref{sec:ancillarity-breakdown}:
Breaking Regressor Ancillarity in linear OLS}
\\
{\em
  Consider joint distributions that share a function $\mu(\xvec)$ as a
  (a.s.) version of their conditional expectation of the response.
  Among these distributions, there exist $\P_1$ and $\P_2$ with
  $\Bbeta(\P_1) \neq \Bbeta(\P_2)$ if and only if $\mu(\xvec)$ is
  nonlinear.
}

\medskip

\noindent See Appendix~\ref{app:non-ancillarity}.
Because $\Bbeta(\P_{1,2})$ depend on $Y$ only through $\mu(\Xvec)$,
the cause of $\Bbeta(\P_1) \neq \Bbeta(\P_2)$ must be a difference in
their regressor distributions.

\begin{figure}[!t]
  \centering
  \includegraphics[width=6in]{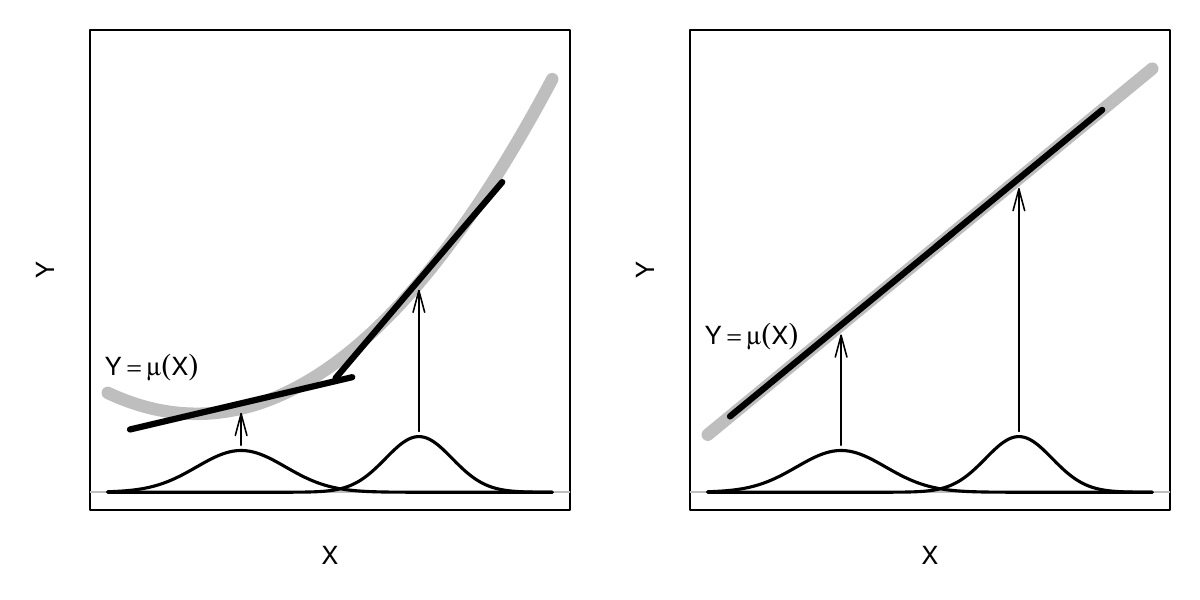}
  \vspace{-.1in}
  \caption{\it Illustration of the dependence of the population OLS
    solution on the marginal distribution of the regressors: The left
    figure shows dependence in the presence of nonlinearity; the right
    figure shows independence in the presence of linearity.}
  \label{fig:nonlinearity-population}
\end{figure}

The proposition is best explained graphically:
Figure~\ref{fig:nonlinearity-population} shows single regressor
scenarios with nonlinear and linear mean functions, respectively, and
the same two regressor distributions.  The two population OLS lines
for the two regressor distributions differ in the nonlinear case and
they are identical in the linear case.\footnote{
See also White (1980a, p.$\,$155f)\nocite{ref:White-1980a}; his $g(Z)+\eps$ is our~$Y$.}

Ancillarity of regressors is sometimes informally explained as the
regressor distribution being independent of, or unaffected by, the
parameters of interest.  From the present point of view where
parameters are not labels for distributions but rather statistical
functionals, this phrasing has things upside down:
\begin{itemize}
\item[] {\em It is not the parameters that affect the regressor
    distribution;  \\ it is the regressor distribution that affects
    the parameters.}
\end{itemize}


\begin{figure}[h!tb]
  \centering
  \includegraphics[height=3.5in]{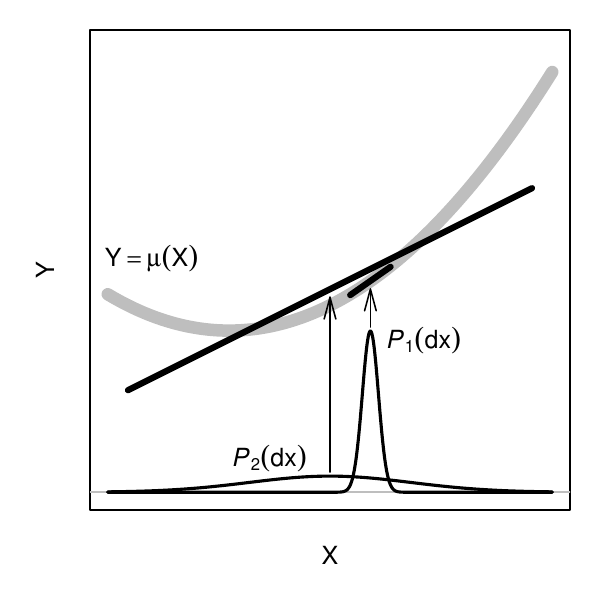}
  \vspace{-.1in}
  \caption{\it Illustration of the interplay between regressors'
    high-density range and nonlinearity: Over the small range of
    $\P_{\!1}$ the nonlinearity is undetectable and immaterial for
    realistic sample sizes, whereas over the extended range of
    $\P_{\!2}$ the nonlinearity is more likely to be detectable and
    relevant.}
  \label{fig:ranges-of-possibilities}
\end{figure}

\subsection{Implications of the Dependence of Slopes on Regressor
  Distributions}

A first practical implication, illustrated by
Figure~\ref{fig:nonlinearity-population}, is that two empirical
studies that use the same regressors, the same response, and the same
model, may yet estimate different parameter values,
$\Bbeta(\P_{\!1}) \!\neq\! \Bbeta(\P_{\!2})$.  This possibility arises
even if the true response surface $\mu(\xvec)$ is identical between
the studies.  The reason is model misspecification and differences
between the regressor distributions in the two studies.  Here is
therefore a potential cause of so-called ``parameter heterogeneity''
in meta-analyses.  ---~The single-regressor situation of
Figure~\ref{fig:nonlinearity-population} gives only an insufficient
impression of the problem because for a single regressor such
differences between regressor distributions are easily detected.  For
multiple regressors the differences take on a multivariate nature and
may become undetectable.

A second practical implication, illustrated by
Figure~\ref{fig:ranges-of-possibilities}, is that misspecification is
a function of the regressor range: Over a narrow range a model has a
better chance of appearing ``correctly specified.''  In the figure the
narrow range of $\P_{\!1}(d\xvec)$ makes the linear approximation
appear very nearly correctly specified, whereas the wide range of
$\P_{\!2}(d\xvec)$ results in gross misspecification.  Again, the
issue gets magnified for larger numbers of regressors where the notion
of ``regressor range'' takes on a multivariate meaning.

Finally, the fact that all models have limited ranges of ``acceptable
approximation'' is a universal issue.  This holds even in those
physical sciences that are based on the most successful theories known
to~us.


\section{The Noise-Nonlinearity Depomposition of OLS Estimates}
\label{sec:estimation}

We turn to estimation from iid~data\footnote{In econometrics, where
  misspecification has been an important topic, the assumption of iid
  data is too limiting; instead, one assumes time series structures.
  See, for example, White (1994)\nocite{ref:White-1994}.}.  We denote
iid~observations from a joint distribution $\PYX$ by
$(Y_i,\Xvec_i\Tr) = (Y_i,1,X_{i,1},...,~X_{i,p})$ ($i=1$,
$2$,~...,~$N$).  We stack them to vectors and matrices as in
Table~\ref{tab:notation}, inserting a constant 1 in the regressors to
accommodate an intercept term.  In particular, $\Xveci\Tr$ is the
$i$'th row and $\Xj$ the $j$'th column of the regressor matrix~$\X$
($i=1,...,N,~ j=0,...,p$).




\begin{table}[h!tb]
\begin{equation*}
  \arraycolsep=1.5pt
  \def\arraystretch{1.7}
  \begin{array}{|lclclll|}
  \hline
  ~~~~\Bbeta    &=& ( \beta_0, \beta_1, ..., \beta_p )\Tr,
                & & & \hbox{    parameter vector } &((p+1) \!\times\! 1)\\
  ~~~~\Y        &=& ( Y_1, ..., Y_N )\Tr,
                & & & \hbox{    response vector } &(N \!\times\! 1)\\
  ~~~~\Xj       &=& ( X_{1,j}, ..., X_{N,j} )\Tr,
                & & & \hbox{    $j$'th regressor vector } &(N \!\times\! 1)\\
  ~~~~\X        &=& [\one,\Xone,...,\Xp]
                &=& \left[\begin{array}{c}
                          ~~\Xvecone\Tr~~ \\ ..... \\ ..... \\ \XvecN\Tr
                          \end{array}\right] ,
                & \begin{array}{l}
                          \hbox{    regressor matrix } \\
                          \hbox{    ~~with intercept }
                    \end{array}
                & (N \!\times\! (p+1))~ \\
  ~~~~\Bmu      &=& ( \mu_1, ..., \mu_N )\Tr,
                & &  \mu_i = \mu(\Xveci) = \E[Y|\Xveci] ,
                & \hbox{    conditional means }
                & (N \!\times\! 1) \\
  ~~~~\Bnonlin  &=& ( \nonlin_1, ..., \nonlin_N )\Tr,
                & &  \nonlin_i = \nonlin(\Xveci) = \mu_i - \Bbeta\Tr \Xveci ,
                & \hbox{    nonlinearities }
                & (N \!\times\! 1) \\
  ~~~~\Bnoise   &=& ( \eps_1, ...,\eps_N )\Tr,
                & &  \eps_i = Y_i - \mu_i ,
                & \hbox{    noise values }
                & (N \!\times\! 1) \\
  ~~~~\Bpopres  &=& ( \popres_1, ..., \popres_N )\Tr,
                & &  \popres_i = \nonlin_i + \eps_i ,
                & \hbox{    population residuals }
                & (N \!\times\! 1) \\
  ~~~~\Bsigma   &=& ( \sig_1, ..., \sig_{\!N} )\Tr,
                & &  \sig_i = \sig(\Xveci) = \V[Y|\Xveci]^{1/2} ,
                & \hbox{    conditional sdevs }
                & (N \!\times\! 1)~~ \\
  ~~~~\Bbetahat &=& ( \betahat_0, \betahat_1, ..., \betahat_p )\Tr
                &=& (\X\Tr\!\X)^{-1} \X\Tr \Y \,,
                & \hbox{    parameter estimates } &((p+1) \!\times\! 1)\\
  ~~~~\r        &=& ( r_1, ..., r_N )\Tr
                &=& \Y - \X \Bbetahat  ,
                & \hbox{    sample residuals }
                & (N \!\times\! 1) \\
  \hline
  \end{array}
\end{equation*}
\caption{Random variable notation for estimation in linear OLS based on iid~observational data.}
\label{tab:notation}
\end{table}



The nonlinearity $\nonlin$, the noise $\eps$, and the population
residuals $\popres$ generate random $N$-vectors when evaluated at all
$N$ observations (again, see Table~\ref{tab:notation}):
\begin{align} \label{eq:defs-vectorized}
  \Bnonlin ~=~ \Bmu\!-\!\X \Bbeta, ~~~~~~~~
  \Bnoise  ~=~ \Y\!-\!\Bmu, ~~~~~~~~
  \Bpopres ~=~ \Y\!-\!\X \Bbeta ~=~ \Bnonlin + \Bnoise.
\end{align}
It is important to distinguish between population and sample
properties: The vectors $\Bpopres$, $\Bnoise$ and $\Bnonlin$ are {\em
  not} orthogonal to the regressor columns $\Xj$ in the sample.
Writing $\langle \cdot, \cdot \rangle$ for the usual Euclidean inner
product on $\Reals^N$, we have in general
\[
  \langle \Bpopres, \Xj \rangle ~\neq~ 0, ~~~~
  \langle \Bnoise,  \Xj \rangle ~\neq~ 0, ~~~~
  \langle \Bnonlin, \Xj \rangle ~\neq~ 0,
\]
even though the associated random variables are orthogonal to $X_j$ in
the population: $\E[\, \popres X_j ] \!=\! 0$,
$\E[\, \eps X_j ] \!=\! 0$,
$\E[\, \nonlin(\Xvec) X_j ] \!=\! 0$, according to~\eqref{eq:orthogonalities}.

\vspace{.05in}

The {\bf OLS estimate} of $\BbetaP$ is as usual
\begin{equation} \label{eq:betahat}
  \arraycolsep=1.5pt
  \def\arraystretch{2.0}
  \begin{array}{|l|}
    \hline
    ~~
    \Bbetahat ~=~ \argmin_\tBbeta \, \|\Y\!-\!\X \tBbeta \|^2
                  ~=~ (\X\Tr\!\X)^{-1} \X\Tr \Y \,.
    \\
    \hline
    \end{array}
\end{equation}
If we write $\Phat$ for the empirical distribution of the observations
$(Y_i,\Xvec_i\Tr)$, then $\Bbetahat = \Bbeta(\Phat)$ is the plug-in
estimate.  Associated is the sample residual vector
$\r = \Y \!-\! \X \Bbetahat$, based on $\Bbetahat$, which is distinct
from the population residual vector $\Bpopres = \Y \!-\! \X \Bbeta$,
based on $\Bbeta = \BbetaP$.

In linear models theory which conditions on (or fixes) $\X$, the
target of estimation is what we may call the ``{\em $\X$-conditional
  parameter}'':
\begin{equation} \label{eq:beta-X}
  \arraycolsep=1.5pt
  \def\arraystretch{2.0}
  \begin{array}{|l|}
    \hline
    ~~
    \BbetaX
    ~~\defn~~ \E[\,\Bbetahat \,|\, \X \,]
    ~=~ \argmin_\Bbeta \, \E[\, \| \Y \!-\! \X \Bbeta \|^2 \,|\, \X \,]
    ~=~ (\X\Tr \X)^{-1} \X\Tr \Bmu .
    ~~
    \\
    \hline
    \end{array}
\end{equation}
In random-$\X$ theory, on the other hand, the target of estimation is
$\BbetaP$, while the $\X$-conditional parameter $\Bbeta(\X)$ is a
random vector.  The vectors $\Bbetahat = \Bbeta(\Phat)$, $\Bbeta(\X)$
and $\BbetaP$ lend themselves to the following telescoping
decomposition:
\begin{equation} \label{eq:Betahat-decomp}
  {\Bbetahat - \BbetaP} ~=~(\Bbetahat - \BbetaX) ~+~ (\BbetaX - \BbetaP) ,
\end{equation}
which in turn reflects the decomposition
$\Bpopres = \Bnoise + \Bnonlin$:

\smallskip

\noindent{\bf Definition and Lemma \ref{sec:estimation}:} {\em Define
  ``Estimation Offsets'' (EOs) as follows:}
\begin{equation} \label{eq:Beta-EOs}
  \arraycolsep=1.5pt
  \def\arraystretch{1.5}
  \begin{array}{|lllll|}
    \hline
 ~~ \textit{Total~EO}         &~~\defn~~&   \Bbetahat  - \BbetaP    &~=~& (\X\Tr\!\X)^{-1} \X\Tr \Bpopres, ~~ \\
 ~~ \textit{Noise~EO}         &~~\defn~~&   \Bbetahat  - \BbetaX    &~=~& (\X\Tr\!\X)^{-1} \X\Tr \Bnoise , ~~ \\
 ~~ \textit{Approximation~EO} &~~\defn~~&   \BbetaX    - \BbetaP    &~=~& (\X\Tr\!\X)^{-1} \X\Tr \Bnonlin. ~~ \\
  \hline
  \end{array}
\end{equation}

\smallskip

\noindent The right sides follow from \eqref{eq:defs-vectorized}, i.e.,
$\Bnoise=\Y\!\!-\!\Bmu$, $\Bnonlin=\Bmu\!-\!\X \Bbeta$,
$\Bpopres=\Y\!\!-\!\X \Bbeta$, and
\begin{equation*}
  \Bbetahat = (\X\Tr\!\X)^{-1} \X\Tr \Y, ~~~~
  \E[\,\Bbetahat\,|\X] = (\X\Tr\!\X)^{-1} \X\Tr \Bmu, ~~~~
  \BbetaP = (\X\Tr\!\X)^{-1} \X\Tr (\X \Bbeta).
\end{equation*}
The first defines $\Bbetahat$, the second uses $\E[\Y|\X] = \Bmu$, and
the third is a tautology.

\smallskip

\noindent{\bf Remark:} One might be tempted to interpret the
approximation EO $\BbetaX \!-\! \BbetaP$ as a bias because it is the
difference of two targets of estimation.  This interpretation is
entirely wrong.  The approximation EO is a random variable when
nonlinearity is present.  It will be seen to contribute not a bias but
a $N^{-1/2}$ order term to the sampling variability of~$\Bbetahat$
(Section~\ref{sec:CLT}).

\begin{figure}[t]
  \centering 
  \includegraphics[width=5.5in]{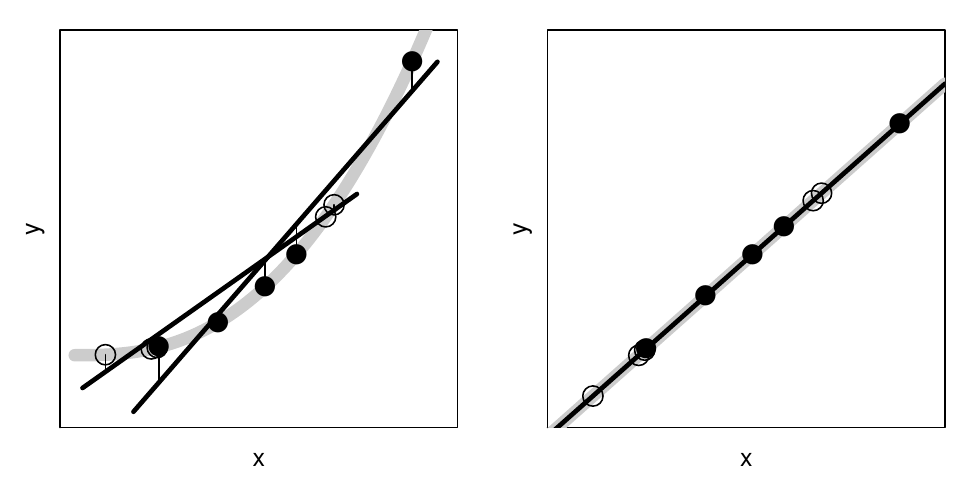}
  \vspace{-.2in}
  \caption{\it Noise-less Response: The filled and the open circles
    represent two ``datasets'' from the same population.  The
    $x$-values are random; the $y$-values are a deterministic function
    of $x$: $y=\mu(x)$ (shown in gray).  \newline Left: The true
    response $\mu(x)$ is nonlinear; the open and the filled circles
    have different OLS lines (shown in black).  Right: The true
    response $\mu(x)$ is linear; the open and the filled circles have
    the same OLS line (black on top of gray). }
  \label{fig:no-error}
\end{figure}


\section{Broken Regressor Ancillarity II:  ~~~~~~~~~~~~~~~~~~~ Nonlinearity  and Random $\X$ Create Sampling Variation}
\label{sec:randomX-plus-nonlinearity}

\subsection{Sampling Variation's Two Sources: Noise AND Nonlinearity}
\label{sec:EV-decomp}

For the $\X$-conditional parameter $\Bbeta(\X)$ to be a non-trivial
random variable, two factors need to be present: (1)~the regressors
$\Xvec$ need to be random and (2)~the nonlinearity $\nonlin(\Xvec)$
must not vanish: $\P[\nonlin(\Xvec) \neq 0] > 0$.  In combination,
these factors conspire to produce sampling variation according to
\eqref{eq:Beta-EOs} which shows the approximation EO to depend
on the random matrix $(\X\Tr\!\X)^{-1} \X\Tr$ and the vector of
nonlinearity values $\Bnonlin$.
\begin{equation} \label{eq:EV-decomp1}
  \V[\,\Bbetahat\,] ~=~ \E[\,\V[\,\Bbetahat \,| \X]]
                    ~+~ \V[\,\E[\,\Bbetahat \,| \X]] \, ,
\end{equation}
where the left side represents the full unconditional variability of
$\Bbetahat$ relevant for statistical inference.  In view of
Lemma~\ref{sec:estimation} this decomposition parallels
$\Bpopres = \Bnoise + \Bnonlin$:
\begin{equation} \label{eq:EV-decomp2}
  \arraycolsep=1.5pt
  \def\arraystretch{1.8}
  \begin{array}{l}
    \,~~~~~~~~~~\V[\,\Bbetahat\,]
    ~=~ \V[\, (\X\Tr\!\X)^{-1} \X\Tr \Bpopres \,] \,,
    \\
    \,~~\E[\,\V[\,\Bbetahat \,| \X]]
    ~=~ \E[\, (\X\Tr\!\X)^{-1} \X\Tr \; \V[\,\Bnoise \,| \X] ~ \X \, (\X\Tr\!\X)^{-1} \,]
    \\
    \begin{array}{|l|}
      \hline
      ~~  \V[\,\E[\,\Bbetahat \,| \X]\,]
      ~=~ \V[\,\BbetaX\,]
      ~=~ \V[\, (\X\Tr\!\X)^{-1} \X\Tr \Bnonlin \,] ~~
      \\     \hline
    \end{array}
  \end{array}
\end{equation}
The center line above the box represents the marginal sampling
variability due to noise combined with randomness in $\X$.  Note that
$\V[\,\Bnoise \,| \X] = D_{\Bsigma^2}$ is the diagonal matrix of noise
variances.  The box shows how the vector of nonlinearities $\Bnonlin$
``conspires'' with the randomness of~$\X$ to generate sampling
variability in~$\BbetaX$.



Intuition for the sampling variability of $\BbetaX$ is best provided
by a graphical illustration.  In order to isolate this effect we
consider a noise-free situation where the response is deterministic
and nonlinear, hence a linear fit is ``misspecified.''  To this end
let $Y \!=\! \mu(\Xvec)$ where $\mu(\cdot)$ is some non-linear
function (that is, $\PYgX$ are point masses $\delta_{\mu(\Xvec)}$),
and hence $\V[\Bbetahat|\X] = \0$ vanishes~a.s.  An example is shown
in the left hand frame of Figure~\ref{fig:no-error} for a single
regressor, with OLS lines fitted to two ``datasets'' consisting of
$N\!=\!5$ regressor values each.  The randomness in the regressors
causes the fitted line to differ between datasets, hence exhibit
sampling variability due to the nonlinearity of the response.  This
effect is absent in the right hand frame of Figure~\ref{fig:no-error}
where the response is linear.\footnote{As in footnote 1, we urge the
  reader to watch a more striking animated illustration of this effect
  by executing the following line of code in an {\em {\bf R}
    Language}~(2008)\nocite{ref:R-2008} interpreter:
  \\
  \centerline{\tt\footnotesize
    source("http://stat.wharton.upenn.edu/\~{}buja/src-conspiracy-animation2.R")
  } }



\subsection{Quandaries of Fixed-$\X$ Theory and the Need for Random-$\X$ Theory}

The fixed-$\X$ approach of linear models theory necessarily assumes
correct specification.  Its only source of sampling variability is the
noise EO $\Bbetahat \!-\! \BbetaX$ arising from the conditional
response distribution, ignoring the approximation EO
$\BbetaX \!-\! \BbetaP$ due to conditioning on $\X$.  A partial remedy
in fixed-$\X$ theory is to rely on diagnostics to detect lack of fit
(misspecification).  We emphasize that diagnostics should be part of
every regression analysis.  In fact, to assist such diagnostics and
make them relevant for correctly sized standard errors, we propose in
Section~\ref{sec:sandwich-adjusted-and-RAV-test} a test to identify
slopes that may have their usual standard errors invalidated by
misspecification.  Furthermore, in Part~II we propose a
misspecification diagnostic for regression parameters.

Data analysts may not stop with negative findings from model
diagnostics and instead continue with data-driven model improvement
by, for example, transforming variables and adding terms to the fitted
equation till the residuals ``look right.''
However, model improvement based on the data can have drawbacks and
limits.  A drawback is that it can invalidate subsequent inferences in
unpredictable ways, as does any data-driven variable selection, formal
or informal (see, e.g., Berk et al.~2013\nocite{ref:Berk-et-al-2013};
Lee et al.~2016\nocite{ref:Lee-et-al-2016}).  A limit is that residual
diagnostics lose power as the number of regressors increases.  This
fact follows from what we may call ``Mammen's dilemma.''  Mammen
(1996)\nocite{ref:Mammen-1996} showed, roughly speaking, that for
models with numerous regressors the residual distribution tends to
look as assumed by the working model, e.g., Gaussian for OLS,
Laplacian for LAD, irrespective of the true error distribution.  For
these reasons, data analysts who diagnose and improve their models
will find themselves torn at some point between hunches of having done
too much of a good thing and missing out on something.

In light of such uncertainties arising from diagnostics and model
improvement, it may be of some comfort that tools are available for
asymptotically correct inference under model misspecification,
including misspecified deterministic responses
($Y \!=\! \mu(\Xvec),~\sigma^2(\Xvec) \!=\! 0)$.  These tools ---
sandwich and $\xy$ bootstrap\footnote{ It needs to pointed out again
  that the residual bootstrap is not assumption-lean.  It requires the
  population residual $\popres$ to be a conventional error term, iid
  across the $N$ observations, implying first and second order correct
  specification ($\nonlin(\Xvec) = 0$ and $\sigma^2(\Xvec) = \sigma^2$
  constant).  The only lean aspect is that the error term no longer
  needs to be Gaussian.  } estimators of standard error --- derive
their justification from central limit theorems (CLTs) to be described
next.


\section{Model-Robust CLTs, Canonically Decomposed}
\label{sec:CLT}

Random-$\X$ CLTs for OLS are standard, and the novel aspect of the
following proposition is in decomposing the overall asymptotic
variance into contributions stemming from the noise EO and the
approximation EO according to~\eqref{eq:Beta-EOs}, thereby providing
an asymptotic analog of the finite-sample decomposition of sampling
variance in Section~\ref{sec:EV-decomp}.

\medskip

\noindent{\bf Proposition~\ref{sec:CLT}:} {\em For linear OLS the
  three EOs follow CLTs:}
\begin{equation} \label{eq:OLS-CLT}
  \arraycolsep=5pt
  \def\arraystretch{1.9}
  \begin{array}{|rcl|}
    \hline
    \sqrt{N} \, (\Bbetahat        - \Bbeta            ) \!\!
    &\stackrel{\Dist}{\longrightarrow}& \!\!
    \Norm\left(\zero, \E[\Xvec \Xvec\Tr]^{-1} \, \E[\, \rmse^2(\Xvec) \Xvec \Xvec\Tr ] \; \E[\Xvec \Xvec\Tr]^{-1} \right)
    \\
    \sqrt{N} \, (\Bbetahat        - \BbetaX) \!\!
    &\stackrel{\Dist}{\longrightarrow}& \!\!
    \Norm\left(\zero, \E[\Xvec \Xvec\Tr]^{-1} \, \E[\, \sig^2(\Xvec) \Xvec \Xvec\Tr ] \; \E[\Xvec \Xvec\Tr]^{-1} \right)
    \\
    \sqrt{N} \, (\BbetaX - \Bbeta     ) \!\!
    &\stackrel{\Dist}{\longrightarrow}& \!\!
    \Norm\left(\zero, \E[\Xvec \Xvec\Tr]^{-1} \, \E[\, \nonlin^2(\Xvec) \Xvec \Xvec\Tr ] \; \E[\Xvec \Xvec\Tr]^{-1} \right)
    \\
    \hline
  \end{array}
\end{equation}

\medskip

\noindent
These three statements once again reflect the decomposition
\eqref{eq:mse}, $\rmse^2(\Xvec) =$ $\sig^2(\Xvec) +$
$\nonlin^2(\Xvec)$.  According to \eqref{eq:sigma2} and \eqref{eq:mse}, $\rmse^2(\Xvec)$ can be replaced by $\popres^2$ and
$\sig^2(\Xvec)$ by~$\eps^2$:
\begin{equation} \label{eq:equivalences}
  \E[\, \rmse^2 (\Xvec) \Xvec \Xvec\Tr ] ~=~ \E[\, \popres^2 \Xvec \Xvec\Tr ],~~~~
  \E[\, \sigma^2(\Xvec) \Xvec \Xvec\Tr ] ~=~ \E[\, \eps^2  \Xvec \Xvec\Tr ].
\end{equation}
The asymptotic variance of linear OLS can therefore be written as
\begin{equation} \label{eq:OLS-AV}
  \arraycolsep=5pt
  \def\arraystretch{1.9}
  \begin{array}{l}
  \AV[\P;\Bbeta] ~~\defn~~
  \E[\Xvec \Xvec\Tr]^{-1} \, \E[\, \popres^2 \Xvec \Xvec\Tr ] \, \E[\Xvec \Xvec\Tr]^{-1} \, .
  \end{array}
\end{equation}
As always, $\Bbeta$ stands for the statistical functional
$\Bbeta = \Bbeta(\P)$ and by implication its plug-in OLS estimator
$\Bbetahat = \Bbeta(\Phat)$.  The formula is the basis for plug-in
that produces the sandwich estimator of standard error
(Section~\ref{sec:plug-in-sandwich}).

Special cases covered by the above proposition are the following:
\begin{itemize} \itemsep 0.5em
\item {\bf First order correct specification:} ~$\nonlin(\Xvec) \overset{\P}{=} 0$.
  The sandwich form is solely due to heteroskedasticity.
\item {\bf Deterministic nonlinear response:} ~$\sig^2(\Xvec)
  \overset{\P}{=} 0$.
  The sandwich form is solely due to the nonlinearity and randomness
  of~$\X$.
\item {\bf First and second order correct specification:} $\nonlin(\Xvec) \!\overset{\P}{=}\! 0$,
  $\sig^2(\Xvec) \!\overset{\P}{=}\! \sig_0^2$.
  The {\em non}-sandwich form is asymptotically valid without Gaussianity:
    $\sqrt{N} \, (\Bbetahat        - \Bbeta            )
    ~\stackrel{\Dist}{\longrightarrow}~
    \Norm\left(\zero,~ \sig_0^2 \, \E[\, \Xvec \Xvec\Tr ]^{-1} \right)$.

\end{itemize}


\section{Sandwich Estimators and the $M$-of-$N$ Bootstrap}
\label{sec:bootstrap-vs-sandwich}

Empirically one observes that standard error estimates obtained from
the $\xy$ bootstrap and from the sandwich estimator are generally
close to each other (Section~\ref{sec:data-example}).  This is
intuitively unsurprising as they both estimate the same asymptotic
variance, that of the first CLT in Proposition~\ref{sec:CLT}.  A
closer connection between them will be described here and established
in generality in Part~II (Buja et
al.~2018\nocite{ref:Buja-et-al-2018b}).\footnote{A third
  assumption-lean method of inference is empirical likelihood.  See
  Owen (2001)\nocite{ref:Owen-2001}.}


\subsection{The Plug-In Sandwich Estimator of Asymptotic Variance}
\label{sec:plug-in-sandwich}

Plug-in estimators of standard error are obtained by substituting the
empirical distribution $\Phat$ for the true $\P$ in formulas for
asymptotic variances.  As the asymptotic variance $\AV[\P;\Bbeta]$ in
\eqref{eq:OLS-AV} is given explicitly and also suitably
continuous in the two arguments, one obtains a consistent estimator by
plugging in $\Phat$ for~$\P$:
\begin{equation} \label{eq:AVhat-functional}
\AVhat\,[\,\Bbeta\,] ~~\defn~~ \AV[\Bbeta, \Phat] ,
~~~~~~~
\SEhat[\betaj] ~~\defn~~ \frac{1}{N^{1/2}} (\AVhat\,[\,\Bbeta\,])_{jj}^{~1/2} .
\end{equation}
[Recall again that $\Bbeta \!=\! \Bbeta(\P)$ stands for the OLS statistical
functional which specializes to its plug-in estimator through $\Bbetahat = \Bbeta(\Phat)$.]
Concretely, one estimates expectations $\E[...]$ with sample means
$\Ehat[...]$, $\Bbeta = \Bbeta(\P)$ with $\Bbetahat = \Bbeta(\Phat)$,
and hence population residuals $\popres^2 = (Y \!-\! \Xvec \Bbeta)^2$
with sample residuals $r_i^2 = (Y_i \!-\! \Xveci \Bbetahat)^2$.
Collecting the latter in a diagonal matrix $\D_\r^2$, one has
\begin{equation*}
  \begin{array}{ccc}
    \Ehat[\,r^2 \, \Xvec \Xvec\Tr]
    =
    \frac{1}{N} \, (\X\Tr \D_\r^2 \, \X) ,
    &~~~~&
    \Ehat[\,\Xvec \Xvec\Tr]
    =
    \frac{1}{N} \, (\X\Tr\!\X).
  \end{array}
\end{equation*}
The sandwich estimator $\AVsand[\,\Bbeta\,] = \AVhat\,[\,\Bbeta\,]$ for
linear OLS in its original form (White 1980a\nocite{ref:White-1980a})
is therefore obtained explicitly as follows:
\begin{equation} \label{eq:OLS-sandwich}
  \begin{array}{rcl}
    \AVsand[\,\Bbeta\,] &\defn&
    \Ehat[\,\Xvec \Xvec\Tr]^{-1} \,
    \Ehat[\, r^2 \Xvec \Xvec\Tr ] \;
    \Ehat[\,\Xvec \Xvec\Tr]^{-1}
    \\[8pt]
    &=& N \,
    (\X\Tr\!\X)^{-1} \,
    (\X\Tr \D_\r^2 \, \X) \,
    (\X\Tr\!\X)^{-1}
  \end{array}
\end{equation}
This is version ``HC'' in MacKinnon and White
(1985)\nocite{ref:MW-1985}.  A modification accounts for the fact that
residuals have smaller variance than noise, calling for a correction
by replacing $1/N^{1/2}$ in \eqref{eq:AVhat-functional} with
$1/(N\!-\!p\!-\!1)^{1/2}$, in analogy to the linear models estimator
(``HC1'' ibid.).  Another modification is to correct individual
residuals for their reduced variance according to
$\V[r_i|\X] = \sig^2 (1-H_{ii})$ under homoskedasticity and ignoring
nonlinearity (``HC2'' ibid.).  Further modifications include a version
based on the jackknife (``HC3'' ibid.) using leave-one-out residuals.
MacKinnon and White (1985)\nocite{ref:MW-1985} also mention that some
forms of sandwich estimators were independently derived by Efron
(1982, p.$\,$18f)\nocite{ref:Efron-1982} using the infinitesimal
jackknife, and by Hinkley (1977)\nocite{ref:Hinkley-1977} using a
``weighted jackknife.''  See Weber (1986)\nocite{ref:Weber-1986} for a
concise comparison in the linear model limited to heteroskedasticity.


\subsection{Sandwich Estimators are Limits of $M$-of-$N$ Bootstrap Estimators}
\label{sec:M-of-N-bootstrap}

An alternative to plug-in is estimating asymptotic variance with the
$\xy$ bootstrap whose justification essentially derives from the
validity of the CLT~.
Conventionally the resample size, here denoted by $M$, is taken to be
the same as the sample size~$N$, but it is useful to distinguish
between these two quantities and allow the resample size $M$ to differ
from $N$, resulting in the ``$M$-of-$N$ bootstrap.''  One
distinguishes
\begin{itemize}
\item $M$-of-$N$ bootstrap resampling {\em with} replacement from
\item $M$-out-of-$N$ subsampling {\em without} replacement.
\end{itemize}
In resampling, $M$ can be any $M \!<\! \infty$; in subsampling, $M$
must satisfy $M \!<\! N$.\footnote{The $M$-of-$N$ bootstrap for
  $M \!\!\ll\!\! N$ ``works'' more often than the conventional
  $N$-of-$N$ bootstrap; see Bickel, G\"otze and van Zwet
  (1997)\nocite{ref:BGvZ-1997} who showed that the favorable
  properties of $M \!\!\ll\!\! N$ subsampling obtained by Politis and
  Romano (1994)\nocite{ref:PR-1994} carry over to the
  $M \!\!\ll\!\! N$ bootstrap.}  To fix notation, denote bootstrap
estimates by $\Bbetaboot_{\!\!\!M} \!=\! \Bbeta(\P_{\!\!M}^*)$, where
$\P_{\!\!M}^*$ is the empirical distribution of bootstrap data
$\{(Y_i^*,\Xveci^*\Tr)\}_{i=1...M}$ drawn iid~from $\Phat_{\!\!N}$.
Bootstrap estimates of asymptotic variance are therefore
\begin{eqnarray} \label{eq:AVboot}
  \AVboot[\,\Bbeta\,] &\defn& M \, \V_{\!\!\!\Phat_{\!\!N}}[\, \Bbetaboot_{\!\!\!M} \,] .
\end{eqnarray}
The connection between bootstrap and sandwich estimates is as follows:





\medskip

\noindent{\bf Proposition~\ref{sec:M-of-N-bootstrap}:}{\em ~The
  sandwich estimator \eqref{eq:OLS-sandwich} is the $M$-of-$N$
  bootstrap estimator \eqref{eq:AVboot} in the limit
  $M \!\!\rightarrow\!\infty$ for a fixed sample of size~$N$.  }

\medskip

See Part II (Buja et al.~2018\nocite{ref:Buja-et-al-2018b}) for full
generality.  Bootstrap approaches may be more flexible than sandwich
approaches because the bootstrap distribution can be used to generate
confidence intervals that are second order correct (see, e.g., Efron
and Tibshirani 1994\nocite{ref:ET-1994}; Hall
1992\nocite{ref:Hall-1992}; McCarthy, Zhang
et.~al.~2016\nocite{ref:McCarthy-Zhang-2016}).


\section{Adjusted Regressors}
\label{sec:adjustment}

This section prepares the ground for two projects: (1)~proposing
meanings of slopes in the presence of nonlinearity
(Section~\ref{sec:ave-slopes}), and (2)~comparing standard errors of
slopes, model-robust versus model-trusting (Section~\ref{sec:AVs}).
The first requires the well-known adjustment formula for slopes in
multiple regression, while the second requires adjustment formulas for
standard errors, both model-trusting and model-robust.  Although the
adjustment formulas are standard, they will be stated explicitly to
fix notation.  [See Appendix \ref{app:adjustment} for more notational
details.]

\begin{itemize} \itemsep 0.5em

\item {\bf Adjustment in Populations:} The {\em population-adjusted
    regressor random variable} $\Xjadj$ is the ``residual'' of the
  population regression of $\XXj$, used as the response, on all other
  regressors.  The response $Y$ can be adjusted similarly, and we may
  denote it by $Y_{\!\bull \, \noj}$ to indicate that $X_j$ is not
  among the adjustors, which is implicit in the adjustment of~$X_j$.
  The multiple regression coefficient~$\betaj = \betaj(\P)$ of the
  population regression of $Y$ on~$\Xvec$ is obtained as the simple
  regression through the origin of $Y_{\! \bull \, \noj}$ or $Y$
  on~$\Xjadj$:
  \begin{equation} \label{eq:pop.adj.formula}
    \betaj
    ~=~ \frac{\E[\,Y_{\! \bull \,\noj} \Xjadj]}{\E[\,\Xjadjsq]}
    ~=~ \frac{\E[\,Y \Xjadj]}{\E[\,\Xjadjsq]}
    ~=~ \frac{\E[\,\mu(\Xvec) \Xjadj]}{\E[\,\Xjadjsq]}
    \,.
  \end{equation}
  The rightmost representation holds because $\Xjadj$ is a function of
  $\Xvec$ only which permits conditioning of $Y$ on $\Xvec$ in the
  numerator.

\item {\bf Adjustment in Samples:} Define the {\em sample-adjusted
    regressor column} $\XBjhadj$ to be the residual vector of the
  sample regression of $\Xj$, used as the response vector, on all
  other regressors.  The response vector $\Y$ can be sample-adjusted
  similarly, and we may denote it by $\YBhadjj$ to indicate that $\Xj$
  is not among the adjustors, which is implicit for $\XBjhadj$.  (Note
  the use of hat notation ``$\,\hbull\,$'' to distinguish it from
  population-based adjustment ``$\bull$.'')  The coefficient estimate
  $\betajhat$ of the multiple regression of $\Y$ on~$\X$ is obtained
  as the simple regression through the origin of $\YBhadjj$ or $\Y$
  on~$\XBjadj$:
  \begin{equation} \label{eq:betahatj}
    \betajhat
    ~=~ \frac{\langle \YBhadjj, \XBjhadj \rangle}{\| \XBjhadj \|^2}
    ~=~ \frac{\langle \Y, \XBjhadj \rangle}{\| \XBjhadj \|^2}
    \,.
  \end{equation}

\end{itemize} [For practice, the patient reader may wrap his/her mind
around the distinction between $\XBjhadj$ and~$\XBjadj$, the latter
being the vector of population-adjusted~$\Xijadj$.  The components of
the former are dependent, those of the latter independent.]



\begin{figure}[t]
  \centering
  \includegraphics[width=5.5in]{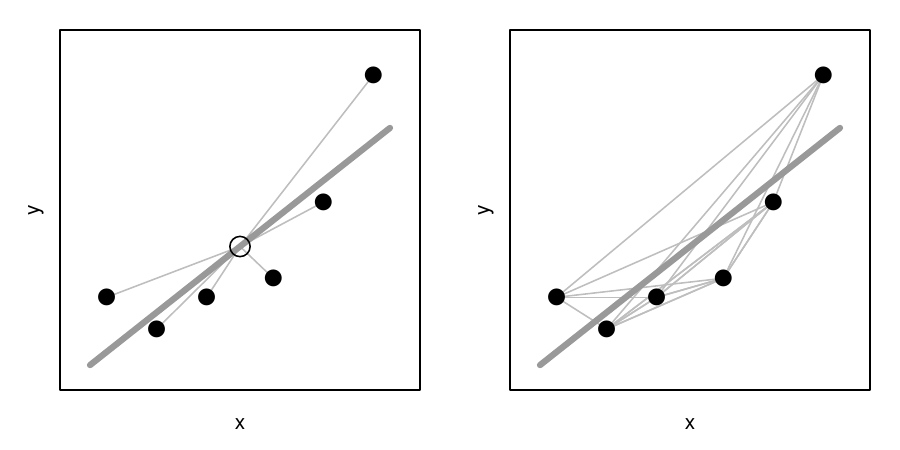}
  \vspace{-.1in}
  \caption{\it Case-wise and pairwise average weighted slopes
    illustrated: Both plots show the same six points (``cases'') as
    well as the OLS line fitted to them (fat gray).  The left hand
    plot shows the case-wise slopes from the mean point (open
    circle) to the six cases, while the right hand plot shows the
    pairwise slopes between all 15 pairs.  In both plots the
    observed slopes are positive with just one exception each,
    supporting the impression that the direction of association is
    positive.}
  \label{fig:ave-slopes}
\end{figure}

\section{Meanings of Slopes in the Presence of Nonlinearity}
\label{sec:ave-slopes}

A first use of regressor adjustment is for proposing meanings of
linear slopes in the presence of nonlinearity, and responding to
Freedman's (2006, p.~302)\nocite{ref:Freedman-2006} objection:
``... it is quite another thing to ignore bias [nonlinearity].  It
remains unclear why applied workers should care about the variance of
an estimator for the wrong parameter.''  Against this view one may
argue that ``flawed'' models are a fact of life.  Flaws such as
nonlinearity can go undetected, or they can be tolerated for
insightful simplification.  A ``parameter'' based on best
approximation is then not intrinsically wrong but in need of a useful
interpretation.

The issue is that, in the presence of nonlinearity, slopes lose their
usual interpretation: $\betaj$ is no longer the average difference in
$Y$ associated with a unit difference in $X_j$ at fixed levels of all
other $X_k$.  Such an interpretation holds for the best approximation
$\Bbeta' \xvec$ but not the conditional mean function $\mu(\xvec)$.
The challenge is to provide an alternative interpretation that remains
valid and intuitive.  As mentioned, a plausible approach is to use
adjusted variables, hence by \eqref{eq:pop.adj.formula} and
\eqref{eq:betahatj} it is sufficient to solve the interpretation
problem for simple regression through the origin.  In a sense to be
made precise, slopes can then be interpreted as weighted averages of
``case-wise'' and ``pairwise'' slopes.
---~To lighten the notational burden, we drop subscripts from adjusted
variables:
\begin{eqnarray*}
  & y \leftarrow \Yadjj    \,,~~~~~~ & x \leftarrow \Xjadj\,,   \,~~~~~~~~
                                      \beta \leftarrow \betaj     \,~~~~  \textrm{for populations,} \\
  & y_i \leftarrow (\YBhadjj)_i \,,~~ & x_i \leftarrow (\XBjhadj)_i\,, ~~~~
                                       \betahat \leftarrow \betajhat ~~~~\textrm{for samples.}
\end{eqnarray*}
By \eqref{eq:pop.adj.formula} and \eqref{eq:betahatj}, the population
slopes and their estimates are, respectively,
\[
  \beta = \frac{E[yx]}{E[x^2]}
  ~~~~~~ \textrm{and} ~~~~~~
  \betahat = \frac{\sum y_i x_i}{\sum x_i^2} .
\]



\noindent Slope interpretation will be based on the following devices:

\begin{itemize} \itemsep .1in

\item {\bf Population parameters} $\beta$ can be represented as weighted
  averages of ...

  \begin{itemize}  \itemsep .2em

  \item {\bf case-wise slopes}: For a random case $(x,y)$ we have
    \begin{equation*}
      \beta = \E[\, w \, b \,] ,
      ~~~\textrm{where}~~~
      b ~\defn~ \frac{y}{x} ,
      ~~~~
      w ~\defn~ \frac{x^2}{\E[\,x^2\,]} .
    \end{equation*}
    Thus $b$ is the case-wise slope through the origin and $w$ its weight.

  \item {\bf pairwise slopes}: For iid cases $(x,y)$ and $(x',y')$ we have
    \begin{equation*}
      \beta ~=~ \E[\, w \, b \,],
      ~~~\textrm{where}~~~
      b ~\defn~ \frac{y-y'}{x-x'} ,
      ~~~~
      w ~\defn~ \frac{(x-x')^2}{\E[\,(x-x')^2\,]} .
    \end{equation*}
    Thus $b$ is the pairwise slope and $w$ its weight.

  \end{itemize}

\item {\bf Sample estimates} $\betahat$ can be represented as weighted
  averages of ...

  \begin{itemize} \itemsep .2em

  \item {\bf case-wise slopes}:
    \begin{equation*} 
      \betahat = \sum_i \, w_i \, b_i \, ,
      ~~~\textrm{where}~~~
      b_i ~\defn~ \frac{y_i}{x_i} ,
      ~~~~
      w_i ~\defn~ \frac{x_i^2}{\sum_{i'} \, x_{i'}^2 \,} .
    \end{equation*}
    Thus $b_i$ are case-wise slopes and $w_i$ their weights.

  \item {\bf pairwise slopes}:
    \begin{equation*} 
      \betahat = \sum_{ik} \, w_{ik} \, b_{ik} \, ,
      ~~~\textrm{where}~~~
      b_{ik} ~\defn~ \frac{y_i - y_k}{x_i - x_k} ,
      ~~~~
      w_{ik} ~\defn~ \frac{(x_i-x_k)^2}{\sum_{i'k'} \, (x_{i'}-x_{k'})^2 \,} .
    \end{equation*}
    Thus $b_{ik}$ are pairwise slopes and $w_{ik}$ their weights
    ($i \neq k$).

  \end{itemize}

\end{itemize}

See Figure~\ref{fig:ave-slopes} for an illustration for samples.
The formulas support the intuition that, even in the presence of
nonlinearity, a linear fit can describe the overall direction of the
association between the response and a regressor after adjustment.

There exist of course examples where no global direction of
association exists, as when $\E[y|x] \!\sim\! x^2$ and the regressor
distribution $\P_{\!\!x}$ is symmetric about $0$.  The association is
local, negative for $x<0$ and positive for $x>0$.  
But if $\E[x]/\SD[x] \gg 0$, the direction of association is positive,
and a linear fit provides an excellent approximation to~$x^2$,
illustrating once again the crucial role of~$\P_{\!\!x}$.

We conclude with a note on the history of the above formulas: Stigler
(2001)\nocite{ref:Stigler-2001} points to Edgeworth, while Berman
(1988)\nocite{ref:Berman-1988} traces them back to an 1841 article by
Jacobi written in Latin.  A generalization based on tuples rather than
pairs of cases was used by Wu (1986)\nocite{ref:Wu-1986} for the
analysis of jackknife and bootstrap procedures (see his Section 3,
Theorem~1).  Gelman and Park (2008)\nocite{ref:GP-2008} also refer to
the representation of OLS slopes as weighted means of pairwise slopes.

\section{Asymptotic Variances --- Proper and Improper}
\label{sec:AVs}

The following prepares the ground for an asymptotic comparison of
model-robust and model-trusting standard errors, one regressor at a
time.


\subsection{Preliminaries: Adjustment Formulas for EOs and Their CLTs:}
\label{sec:adjustment-CLTs}

The vectorized formulas for estimation offsets
\eqref{eq:Betahat-decomp} can be written componentwise using
adjustment as follows:
\begin{equation} \label{eq:EOs-adj}
  \arraycolsep=3pt
  \def\arraystretch{2.5}
  \begin{array}{|lrcl|}
    \hline
    \textit{~Total~EO}:            & \betajhat  - \betaj
      &=& \DS \frac{\langle \XBjhadj, \Bpopres \rangle}{\|\XBjhadj\|^2},~
    \\[2pt]
    \textit{~Noise~EO}:            & \betajhat   - \betaj(\X)
      &=& \DS \frac{\langle \XBjhadj, \Bnoise    \rangle}{\|\XBjhadj\|^2},~
    \\[2pt]
    \textit{~Approximation~EO}:~~~ & \betaj(\X) - \betaj
      &=& \DS \frac{\langle \XBjhadj, \Bnonlin \rangle}{\|\XBjhadj\|^2}.~
    \\[6pt]
    \hline
  \end{array}
\end{equation}
To see these identities directly, note the following, in addition to
\eqref{eq:betahatj}:
$\E[\betajhat|\X] = \langle \Bmu, \XBjhadj \rangle / \|\XBjhadj\|^2$
and $\betaj = \langle \X \Bbeta, \XBjhadj \rangle/ \|\XBjhadj\|^2$,
the latter due to
$\langle \XBjhadj, \X_k \rangle = \delta_{jk} \|\XBjhadj\|^2$.
Finally use $\Bpopres = \Y\!-\!\X \Bbeta$,
$\Bnonlin = \Bmu \!-\! \X\Bbeta$ and
$\Bnoise = \Y\!-\!\Bmu$.~~$\square$

From \eqref{eq:EOs-adj}, asymptotic normality of the
coefficient-specific EOs can be separately expressed using population
adjustment:

\medskip

\noindent
\begin{samepage}
{\bf Corollary \ref{sec:adjustment-CLTs}:} 
\begin{equation*}
  \arraycolsep=3pt
  \def\arraystretch{2.7}
  \begin{array}{|ccccc|}
    \hline
    ~~N^{1/2} (\betajhat - \betaj)
    & \stackrel{\Dist}{\longrightarrow} &
    \DS \Norm\left(0, \frac{\E[\,\rmse^2(\Xvec) \Xjadjsq]}{\E[\,\Xjadjsq]^2} \right)
    &=&
    \DS \Norm\left(0, \frac{\E[\,\popres^2 \Xjadjsq]}{\E[\,\Xjadjsq]^2} \right) ~
    \\[16pt]
    ~~N^{1/2} (\betajhat - \betaj(\X))
    & \stackrel{\Dist}{\longrightarrow} &
    \DS \Norm\left(0, \frac{\E[\,\sigma^2(\Xvec) \Xjadjsq]}{\E[\,\Xjadjsq]^2} \right)
    &=&
    \DS \Norm\left(0, \frac{\E[\,\eps^2 \Xjadjsq]}{\E[\,\Xjadjsq]^2} \right)
    \\[16pt]
    ~N^{1/2} (\betaj(\X) - \betaj)
    & \stackrel{\Dist}{\longrightarrow} &
    \DS \Norm\left(0, \frac{\E[\,\nonlin^2(\Xvec) \Xjadjsq]}{\E[\,\Xjadjsq]^2} \right)
    & &
    \\[6pt]
    \hline
  \end{array}
\end{equation*}
\end{samepage}

\noindent
The equalities on the right side in the first and second case are
based on \eqref{eq:equivalences}.  The first CLT in its right side
form is useful for plug-in estimation of asymptotic variance, one
slope at a time.
The sandwich form of matrices has been reduced to ratios where
numerators correspond to the ``meat'' and squared denominators to the
``breads.''

\smallskip


\subsection{Model-Robust Asymptotic Variances in Terms of Adjusted Regressors:}
\label{sec:AVproper}

The CLTs of Corollary~\ref{sec:adjustment-CLTs} contain three
asymptotic variances of the same form with arguments $\rmse^2(\Xvec)$,
$\sig^2(\Xvec)$ and $\nonlin^2(\Xvec)$.  We will use $\rmse^2(\Xvec)$
in the following definition for the overall asymptotic variance, but
by substituting $\sig^2(\Xvec)$ or $\nonlin^2(\Xvec)$ for
$\rmse^2(\Xvec)$ one obtains terms that can be interpreted as
components of the overall asymptotic variance or else as asymptotic
variances in the absence of nonlinearity or absence of noise.

\medskip

\noindent{\bf Definition \ref{sec:AVproper}:} 
{\em Proper Asymptotic Variance.}
\begin{equation*}
  \def\arraystretch{1.1}
  \begin{array}{|c|}
    \hline \vspace{-.13in}
    \\
    \AVlean[\betaj;\rmse^2]  ~~\defn~~  \DS \frac{\E[\,\rmse^2(\Xvec) \Xjadjsq]}{\E[\,\Xjadjsq]^2} .
    \\[1em]
    \hline
  \end{array}
\end{equation*}

\vspace{.05in}



\noindent
From \eqref{eq:mse}, $\rmse^2(\Xvec) \!=\! \sig^2(\Xvec) \!+\! \nonlin^2(\Xvec)$,
one obtains
\begin{equation*}
  \AVlean[\betaj;\rmse^2] ~=~
  \AVlean[\betaj;\sig^2] + \AVlean[\betaj;\nonlin^2] .
\end{equation*}
The subscript ``$\lean$'' refers to validity in the assumption-lean
model-robust framework.  This proper asymptotic variance will be
compared to the potentially improper asymptotic variance of
model-trusting linear models theory (Section~\ref{sec:RAV}).

\smallskip


\subsection{Model-Trusting Asymptotic Variances in Terms of Adjusted Regressors}
\label{sec:AVimproper}

The goal is to provide an asymptotic limit for the usual
model-trusting standard error estimate of linear models theory in the
model-robust framework.  To this end we need the model-robust limit of
the usual estimate of the noise variance,
$\sigmahat^2 = \|\Y -\X \Bbetahat\|^2 / (N\!-\!p\!-\!1)$:
\begin{equation*}
  \sigmahat^2  ~~\overset{\P}{\longrightarrow}~~
  \E[\, \popres^2 \,] ~=~
  \E[\, \rmse^2(\Xvec) \,] ~=~
  \E[\, \sig^2(\Xvec) \,] + \E[\, \nonlin^2(\Xvec) \,] ,~~~~~~
  N \rightarrow \infty.
\end{equation*}
Thus the model-robust limit of $\sigmahat^2$ is the average
conditional MSE of $Y$, which again decomposes according to
$\rmse^2(\Xvec) \!=\! \sig^2(\Xvec) \!+\! \nonlin^2(\Xvec)$.

Squared standard error estimates are, in matrix and adjustment form,
\begin{equation}  \label{eq:sigmahat}
  \Vhat_\Vlin[\,\Bbetahat\,] ~=~ \sigmahat^2 \, (\X\Tr\!\X)^{-1}
  ,~~~~~~~~
  \SEhat^2_\lin[\,\betahat] ~=~ \frac{\sigmahat^2}{\|\XBjhadj\|^2}.
\end{equation}
Their assumption-lean scaled limits are
\begin{equation*} 
  N\, \Vhat_\Vlin[\,\Bbetahat\,]
  ~\overset{\P}{\longrightarrow}~
  \E[\, \rmse^2(\Xvec) \,] ~ \E[\, \Xvec \Xvec\Tr \,]^{-1}
  ,~~~~~
  N\, \SEhat^2_\lin[\,\betajhat]
  ~\overset{\P}{\longrightarrow}~
  \frac{ \E[\, \rmse^2(\Xvec) \,] }{ \E[\, X_{j \bull}^2 \,] } .
\end{equation*}

\medskip

\noindent{\bf Definition \ref{sec:AVimproper}}
{\em Improper Asymptotic Variance.}
\begin{equation*}
  \def\arraystretch{1.1}
  \begin{array}{|c|}
    \hline \vspace{-.13in}
    \\
    \AVlin[\betaj;\rmse^2]  ~~\defn~~ \DS \frac{\E[\,\rmse^2(\Xvec)]}{\E[\,\Xjadjsq]} .
    \\[1em]
    \hline
  \end{array}
\end{equation*}



\noindent
This decomposes once again according to
$\rmse^2(\Xvec) \!=\! \sig^2(\Xvec) \!+\! \nonlin^2(\Xvec)$:
\begin{equation*}
  \AVlin[\betaj;\rmse^2] ~=~
  \AVlin[\betaj;\sig^2] + \AVlin[\betaj;\nonlin^2] .
\end{equation*}
The subscript ~$\lin$ refers to validity of this asymptotic variance
under the assumption-loaded model-trusting framework of linear models
theory.

\smallskip


\subsection{$\RAV$ ---  Ratio of Proper and Improper Asymptotic Variances}
\label{sec:RAV}

To examine the discrepancies between proper and improper asymptotic
variances we form their ratio, which results in the following elegant
functional of the conditional MSE and the squared adjusted regressor:

\medskip

\noindent{\bf Definition \ref{sec:RAV}:}{\em~~Ratio of Asymptotic Variances ($\RAV\!$), Proper/Improper.}
\\
\begin{equation*}
  \def\arraystretch{2.5}
  \begin{array}{|ccccc|}
    \hline
    \RAV[\betaj,\rmse^2]
    &\defn&
    \DS \frac{\AVlean[\betaj,\rmse^2]}{\AVlin[\betaj,\rmse^2]}
    &=&
    \DS \frac{\E[\rmse^2(\Xvec) \Xjadjsq]}{\E[\rmse^2(\Xvec)]\,\E[\Xjadjsq]} . \\[.9em]
    \hline
  \end{array}
\end{equation*}

\noindent
In order to examine the effect of heteroskedasticities and
nonlinearities on the discrepancies separately, one can also define
$\RAV[\betaj,\sig^2]$ and $\RAV[\betaj,\nonlin^2]$.  By the
decomposition lemma in Appendix~\ref{app:RAV-decomposition},
$\RAV[\betaj,\rmse^2]$ is a weighted mixture of these two terms.
---~Belaboring the obvious, the interpretation of the $\RAV$ is:

\vspace{-1em}

\begin{equation*}
  \textrm{If~~} \RAV[\betaj,\rmse^2]
  \left\{
  \begin{array}{c}  > 1 \\ = 1 \\ < 1  \end{array}
  \right\},
  \textrm{then~} \SEhat_\lin[\betajhat]
  \textrm{~is~asymptotically~}
  \left\{
  \begin{array}{l}
    \textrm{too~small} \\
    \textrm{correct} \\
    \textrm{too~large}
  \end{array}
  \right\}
  .
\end{equation*}



We will later have use for the following sufficient condition
for~$\RAV\!=\!1$.  It says essentially that when the population
residual $\popres$ is a traditional error term, then the usual
standard error of linear models theory is asymptotically correct.  The
condition is equivalent to first and second order correct
specification, including linearity and homoskedasticity but not
Gaussianity.

\medskip


\noindent{\bf Lemma \ref{sec:RAV}:}~{\em Sufficient conditions
  for $\RAV[\betaj,\rmse^2] = 1$ are the following:
  \begin{itemize}
  \item[(a)] $\rmse^2(\Xvec) = \rmse_0^2$ is constant.
  \item[(b)] $\popres^2$ and $\Xjadj^2$ are independent.
  \end{itemize}
}


\noindent {\bf Proof}: $(a)$ is immediate from
Definition~\ref{sec:RAV}.  $(b)$~The numerator of
$\RAV[\betaj,\rmse^2]$ is
$\E[\rmse^2(\Xvec) \Xjadj^2] \!=\! \E[\,\popres^2 \Xjadj^2] \!=\!
\E[\,\popres^2] \, \E[\Xjadj^2]$, hence equals the denominator.
$\square$

\medskip

The ratio $\RAV[\betaj,\rmse^2]$ is the inner product between the
random variables
\[
  \frac{\rmse^2(\Xvec)}{\E[\rmse^2(\Xvec)]}~~~~
  \mbox{and}~~~~
  \frac{\Xjadjsq}{\E[\Xjadjsq]} .
\]
It is {\em not} a correlation as both $\rmse^2(\Xvec)$ and $\Xjadjsq$
are $L_1$-normalized; a non-centered correlation would require
$L_2$-normalization with denominators $\E[\rmse^4(\Xvec)]^{1/2}$ and
$\E[{X_{\!j \bull}^{\,4}}]^{1/2}$, respectively.  Its upper bound is
obviously not $+1$ but rather~$\infty$:

\smallskip



\subsection{The Range of $\RAV$}
\label{sec:range-of-RAV}

The analysis of the $\RAV$ is simplified by conditioning
$\rmse^2(\Xvec)$ on $\Xjadjsq$:

\medskip

\noindent{\bf Definition and Lemma \ref{sec:range-of-RAV}:}~{\em Letting
  \[
  \rmse_j^2(\Xjadjsq) ~~\defn~~ \E[\rmse^2(\Xvec) \,|\,\Xjadjsq ],
  \]
we have:}
\begin{equation*}
  \RAV[\betaj,\rmse^2] ~=~ \RAV[\betaj,\rmse_j^2] .
\end{equation*}

\medskip

\noindent
Thus the analysis of the $\RAV$ is reduced to single squared adjusted
regressors $\Xjadjsq$.  This fact lends itself to simple case studies
and graphical illustrations.

\smallskip

Next we describe the extremes of the $\RAV$ over scenarios of
$\rmse^2(\Xvec)$ or, by Lemma~\ref{sec:range-of-RAV}, of
$\rmse_j^2(\Xjadjsq)$.

\medskip




\noindent{\bf Proposition~\ref{sec:range-of-RAV}:} {\em
If $\E[\Xjadjsq] < \infty$ and $\Xjadjsq$ has unbounded support, then
\vspace{-.5em}
\begin{equation*}
  \sup_{\rmse_j^2} \RAV[\betaj,\rmse_j^2] ~=~ \infty.
\end{equation*}
\vspace{-.5em}
If $\E[\Xjadjsq] < \infty$ and $\Xjadjsq$ has 0 in its support, then
\begin{equation*}
  \inf_{\rmse_j^2} \RAV[\betaj,\rmse_j^2] ~=~ 0.
\end{equation*}
\vspace{-.5em}
}

\begin{figure}[t]
  \centering
  \includegraphics[width=3.0in]{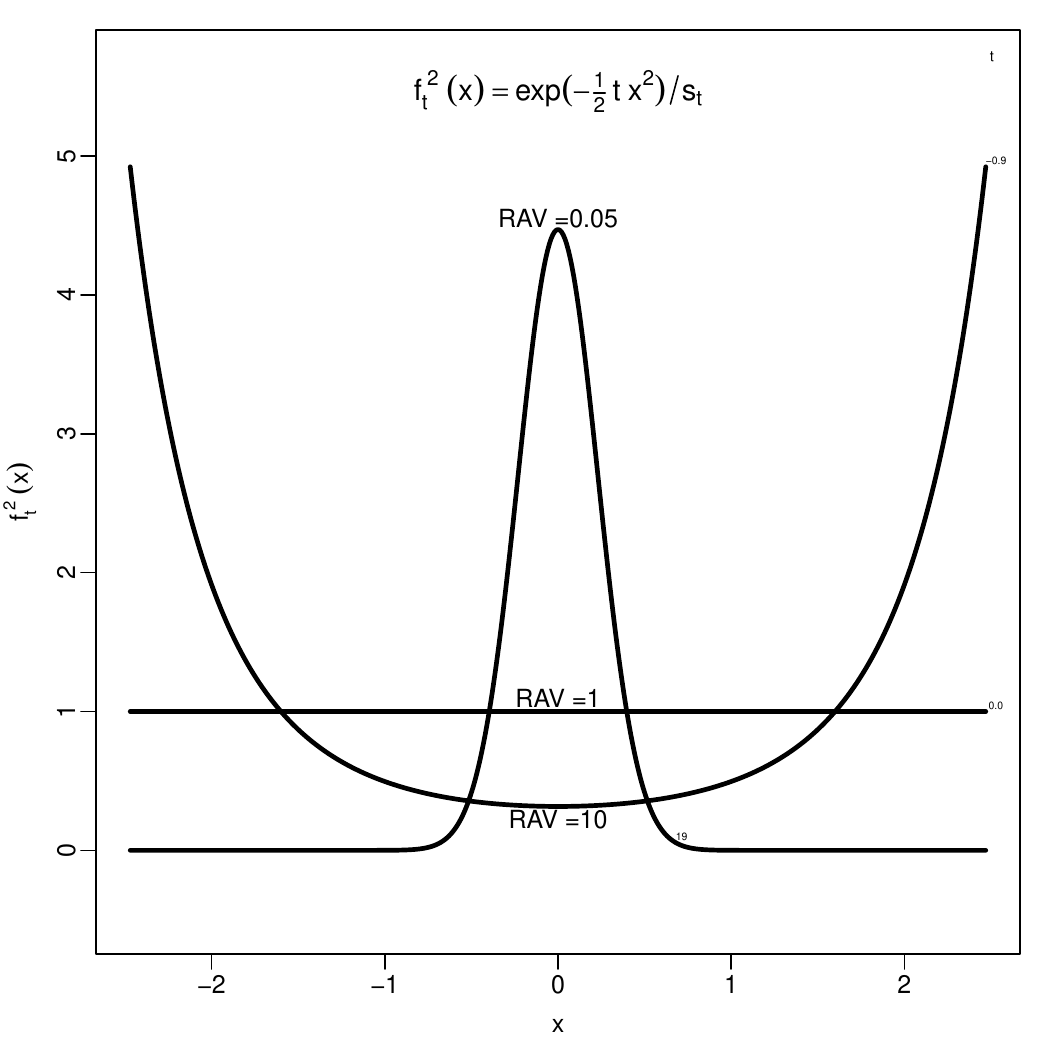}
\vspace{-.1in}
\caption{\it A family of functions $f_t^2(x)$ that can be interpreted
  as conditional MSEs $\rmse_j^2(\Xjadjsq)$, heteroskedasticities
  $\sigma_j^2(\Xjadjsq)$ or squared nonlinearities
  $\nonlin_j^2(\Xjadjsq)$ (shown as functions of $x = \Xjadj$ rather
  than $\Xjadjsq$): The family interpolates $\RAV$ from $0$ to
  $\infty$ for $x = \Xjadj \sim N(0,1)$.  The three solid black curves
  show $f_t^2(x)$ that result in RAV=0.05, 1, and 10.  (See Appendix
  \ref{app:examples-for-sig-and-eta} for details.)  \newline
  $\RAV = \infty$ is approached as $f_t^2(x)$ bends ever more strongly
  in the tails of the $x$-distribution.  \newline $\RAV = 0$ is
  approached by an ever stronger spike in the center of the
  $x$-distribution.  }
\label{fig:RAV-f}
\end{figure}

\begin{figure}[t]
  \centering
  \includegraphics[width=6in]{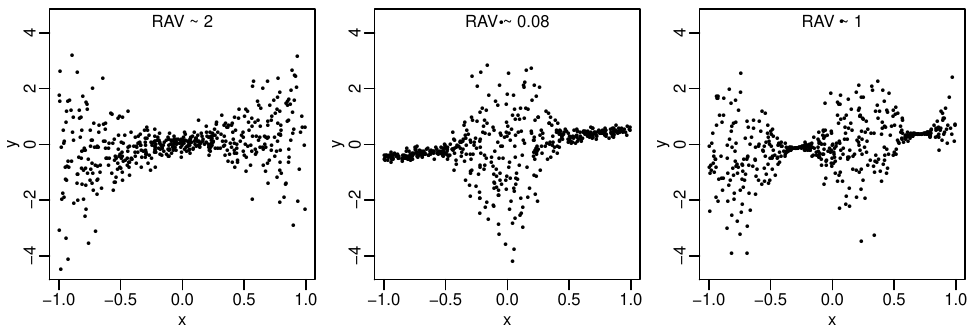}
  \vspace{-.1in}
  \caption{\it The effect of heteroskedasticity on the sampling
    variability of slope estimates: How does the treatment of the
    heteroskedasticities as homoskedastic affect statistical
    inference?  \newline Left: High noise variance in the tails of the
    regressor distribution elevates the true sampling variability of
    the slope estimate above the usual standard error:
    $\RAV[\betaj,\sig^2] > 1$.  \newline Center: High noise variance
    near the center of the regressor distribution lowers the true
    sampling variability of the slope estimate below the usual
    standard error: $\RAV[\betaj,\sig^2] < 1$.  \newline Right: The
    noise variance oscillates in such a way that the usual standard
    error is coincidentally correct ($\RAV[\betaj,\sig^2] = 1$). }
  \label{fig:RAV-3heteros}
\end{figure}

\begin{figure}[t]
  \centering
  \includegraphics[width=6in]{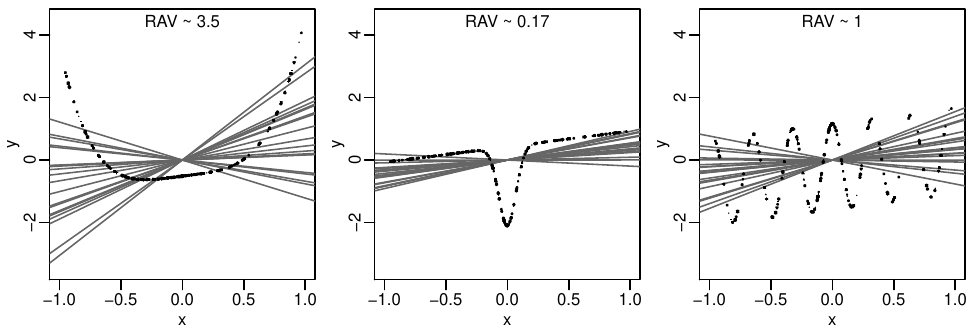}
  \caption{\it The effect of nonlinearities on the sampling
    variability of slope estimates: The three plots show three
    different noise-free nonlinearities; each plot shows for one
    nonlinearity 20 overplotted datasets of size $N=10$ and their
    fitted lines through the origin.  The question is how the
    misinterpretation of the nonlinearities as homoskedastic random
    errors affects statistical inference.  \newline Left: Strong
    nonlinearity in the tails of the regressor distribution elevates
    the true sampling variability of the slope estimate above the
    usual standard error ($\RAV[\betaj,\nonlin^2] > 1$).  \newline
    Center: Strong nonlinearity near the center of the regressor
    distribution lowers the true sampling variability of the slope
    estimate below the usual standard error
    ($\RAV[\betaj,\nonlin^2] < 1$).  \newline Right: An oscillating
    nonlinearity mimics homoskedastic random error to make the usual
    standard error coincidentally correct
    ($\RAV[\betaj,\nonlin^2] = 1$).  \newline Caveat: These are
    cartoons illustrating potential causes of standard error
    discrepancies.  Nonlinearities may not be detectable in actual
    data in the presence of noise and other regressors.  }
  \label{fig:RAV-3nonlins}
\end{figure}

\noindent
Thus, when the adjusted regressor distribution is unbounded, the usual
standard error can be too small to any degree.  Conversely, if the
adjusted regressor is not bounded away from zero, it can be too large
to any degree.

What shapes of $\rmse_j^2(\Xjadjsq)$ approximate these extremes?  The
answer can be gleaned from Figure~\ref{fig:RAV-f} which illustrates
the proposition for normally distributed $\Xjadj$: If {\em
  nonlinearities and/or heteroskedasticities blow up~...}
\begin{itemize}
\item in the {\em tails} of the $\Xjadj$ distribution, then $\RAV$
  takes on {\em large} values;
\item in the {\em center} of the $\Xjadj$ distribution, then $\RAV$
  takes on {\em small} values.
\end{itemize}
The proof in Appendix~\ref{app:extreme.RAV} bears this out.  As the
main concern is with usual standard errors that are too small,
$\RAV \!>\! 1$, the proposition indicates that $\Xjadj$-distributions
with bounded support enjoy some protection from the worst case.

\smallskip


\subsection{Illustration of Factors that Drive the $\RAV$}
\label{sec:range-of-RAV-simulation-examples}

We further analyze the $\RAV\!$ in terms of the constituents
of~$\rmse_j^2(\Xjadjsq)$, conditional variance and squared
nonlinearity, as functions of~$\Xjadjsq$:
\begin{equation} \label{eq:drivers-of-RAV}
\sig_j^2(\Xjadjsq) = \E[\sig^2(\Xvec)|\Xjadjsq]
~~~~\textrm{and}~~~~
\nonlin_j^2(\Xjadjsq) = \E[\nonlin^2(\Xvec)|\Xjadjsq] .
\end{equation}
For qualitative insights into the drivers of the $\RAV\!\!$, we
translate \eqref{eq:drivers-of-RAV} to concrete data scenarios.
Figure~\ref{fig:RAV-3heteros} shows three noise scenarios and
Figure~\ref{fig:RAV-3nonlins} three nonlinearity scenarios.  The
illustrated effects will both be present to degrees in real data.
Their combined effect is described by a decomposition lemma in
Appendix \ref{app:RAV-decomposition}: $\RAV[\betaj,\rmse_j^2]$ is a
weighted mixture of $\RAV[\betaj,\sig_j^2]$ and
$\RAV[\betaj,\nonlin_j^2]$.  Therefore:
\begin{itemize} \itemsep .3em
\item Heteroskedasticities with large $\sig_j^2(\Xjadjsq)$ in the
  tails of $\Xjadj^2$ produce an upward contribution to
  $\RAV[\betaj,\rmse_j^2]$; heteroskedasticities with large
  $\sig_j^2(\Xjadjsq)$ near $\Xjadjsq = 0$ imply a downward
  contribution to $\RAV[\betaj,\rmse_j^2]$.
\item Nonlinearities with large average values $\nonlin_j^2(\Xjadjsq)$
  in the tails of $\Xjadjsq$ imply an upward contribution to
  $\RAV[\betaj,\rmse_j^2]$; nonlinearities with large
  $\nonlin_j^2(\Xjadjsq)$ concentrated near $\Xjadjsq = 0$ imply a
  downward contribution to~$\RAV[\betaj,\rmse_j^2]$.
\end{itemize}
These facts also suggest that large values $\RAV \!\!>\! 1$ should
occur more often than small values $\RAV \!\!<\! 1$ because large
conditional variances as well as nonlinearities are often more
pronounced in the extremes of regressor distributions, not their
centers.  This is most natural for nonlinearities which are often
convex or concave.  Also, it follows from the $\RAV\!$ decomposition
lemma (Appendix \ref{app:RAV-decomposition}) that
either of $\RAV[\betaj,\sigma_j^2]$ or $\RAV[\betaj,\nonlin_j^2]$ is
able to single-handedly pull $\RAV[\betaj,\rmse_j^2]$ to $+\infty$,
whereas both have to be close to zero to pull $\RAV[\betaj,\rmse_j^2]$
toward zero.  These heuristics support the observation that in
practice $\SEhat_\lin$ is more often too small than too large compared
to the asymptotically correct~$\SEhat_\sand$.


\section{Sandwich Estimators in Adjusted Form and a $\RAV$ Test}
\label{sec:sandwich-adjusted-and-RAV-test}


The goal here is to write the $\RAV$ in adjustment form and estimate
it with plug-in for use as a test statistic to decide whether the
usual standard error is adequate.  We will obtain one test per
regressor.  The proposal is related to the class of ``misspecification
tests'' for which there exists a literature starting with Hausman
(1978)\nocite{ref:Hausman-1978} and continuing with White (1980a,b;
1981;
1982)\nocite{ref:White-1980a,ref:White-1980b,ref:White-1981,ref:White-1982}
and others.  These tests are largely global rather than
coefficient-specific, which ours is.  The test proposed here has
similarities to White's (1982, Section 4)\nocite{ref:White-1982}
``information matrix test'' which compares two types of information
matrices globally, while we compare two types of standard errors, one
coefficient at a time.


\smallskip


\subsection{Sandwich Estimators in Adjustment Form and the $\RAVjhat$ Test Statistic}
\label{sec:RAV-test-statistic}

The adjustment versions of the asymptotic variances in the CLTs of
Corollary~\ref{sec:adjustment-CLTs} can be used to rewrite the
sandwich estimator by replacing expectations $\E[...]$ with means
$\Ehat[...]$, $\Bbeta$ with $\Bbetahat$, $\Xjadj$ with $\XBjhadj$, and
rescaling by~$N$:
\begin{equation} \label{eq:sandwich-AV-adj}
  \SEhat_{\sand}[\betajhat]^2
  ~=~
  \frac{1}{N} \frac{\Ehat[\,(Y - \Xvec\Tr \Bbetahat)^2 \Xjhadj^2]}{\Ehat[\,\Xjhadj^2]^{\,2}}
  ~=~
  \frac{\langle \r^2, \XBjhadj^2 \rangle}{\|\XBjhadj\|^4}    
  .
\end{equation}
The squaring of $N$-vectors is meant to be coordinate-wise.  Formula
\eqref{eq:sandwich-AV-adj} is algebraically equivalent to the diagonal
elements of~\eqref{eq:OLS-sandwich}.

To match the raw plug-in form of the sandwich estimator
\eqref{eq:sandwich-AV-adj}, we use the plug-in version of the standard
error estimator 
of linear models theory, the only difference being division by $N$
rather than $N\!-\!p\!-\!1$:
\begin{equation} \label{eq:SEhatlin}
  \SEhat_{\lin}[\betajhat]^2
  ~=~
  \frac{1}{N} \frac{ \Ehat[ (Y -\Xvec\Tr \Bbetahat)^2 ] }{ \Ehat[ \Xjhadj^2 ] }
  ~=~
  \frac{1}{N} \frac{ \| \r \|^2 }{ \|\XBjhadj\|^2 }   
  \, ,
\end{equation}
Thus the plug-in estimate of $\RAV[\betaj,\rmse^2]$ is
\begin{equation} \label{eq:RAV-est}
  \RAVjhat
  ~~\defn~~
  \frac{ \Ehat[\,(Y-\Xvec\Tr \Bbetahat)^2 \Xjhadj^2 \,] }
       { \Ehat[\,(Y-\Xvec\Tr \Bbetahat)^2\,] \, \Ehat[\,\Xjhadj^2\,] }
  ~=~
  N \, \frac{ \langle \r^2, \XBjhadj^2 \rangle }  
            { \| \r \|^2 \; \| \XBjhadj \|^2 }
  \,.
\end{equation}
This is the proposed test statistic.  Analogous to the
population-level $\RAV[\betaj,\rmse^2]$, the sample-level
$\RAVjhat$ responds to associations between squared residuals and
squared adjusted regressors.

\smallskip


\subsection{The Asymptotic Null Distribution of the $\RAV$ Test Statistic}
\label{sec:RAV-test-asymptotic}

Here is an asymptotic result that would be expected to yield
approximate inference under a null hypothesis that
implies~$\RAV[\betaj,\rmse^2]=1$ based on Lemma~\ref{sec:RAV}:

\bigskip

\noindent{\bf Proposition~\ref{sec:RAV-test-asymptotic}:}
~{\em Under the null hypothesis $H_0$ that the population residuals
  $\popres$ and the adjusted regressor $\Xjadj$ are independent, it holds: }
\begin{equation}  \label{eq:RAVhat-asymp-normality}
    N^{1/2} \, ( \RAVjhat - 1 )
    ~ \overset{\Dist}{\longrightarrow} ~
    \Norm\left( 0,\, \frac{\E[\,\popres^4]}{\E[\,\popres^2]^2}  \,
                     \frac{\E[\Xjadj^4]}{\E[\Xjadjsq]^2}
                     \,-\, 1 )
         \right) .
       \end{equation}

\noindent As always we ignore technical assumptions.  A proof
outline is in Appendix~\ref{app:RAV-test}.

The asymptotic variance of $\RAVjhat$ under $H_0$ is driven by the
standardized fourth moments or the kurtoses (= same$\,-\,$3) of
$\popres$ and $\Xjadj$.  Some observations:
\begin{enumerate}
\item The larger the kurtosis of population residuals $\popres$ and/or
  adjusted regressors $\Xjadj$, the less likely is detection of first
  and second order model misspecification resulting in standard error
  discrepancies.
\item As standardized fourth moments are always $\ge 1$ by Jensen's
  inequality, the asymptotic variance is $\ge 0$, as it should be.
  The asymptotic variance vanishes iff the minimal standardized fourth
  moment is $+1$ for both $\popres$ and $\Xjadj$, hence both have
  symmetric two-point distributions (as both are centered).  For such
  $\Xjadj$ it holds $\RAV[\betaj,\rmse^2] \!=\! 1$ by
  Proposition~\ref{app:extreme.RAV} in the appendix.
\item A test of the stronger $H_0$ that includes normality of
  $\popres$ is obtained by setting $\E[\popres^4]/\E[\popres^2]^2 $
  $= 3$ rather than estimating it.  The result, however, is an overly
  sensitive non-normality test much of the time, which does not seem
  useful as non-normality can be diagnosed and tested by other means.
\end{enumerate}

\begin{table}
\resizebox{\textwidth}{!}{\tt
\begin{tabular}{l|rrrrrrr}
                  &$\betajhat$ & $\SE_\lin$ & $\SE_\sand$ & $\RAVjhat$ & \textrm{2.5\% Perm.} & \textrm{97.5\% Perm.} \\
\hline                                                                     
(Intercept)       &   0.760 & 22.767  &  16.209 & 0.495~  &  0.354 &  3.182 \\
MedianIncome (\$K)&  -0.183 &  0.187  &   0.108 & 0.318~  &  0.274 &  5.059 \\
PercVacant        &   4.629 &  0.901  &   1.363 & 2.071~  &  0.303 &  3.823 \\
PercMinority      &   0.123 &  0.176  &   0.164 & 0.860~  &  0.403 &  2.238 \\
PercResidential   &  -0.050 &  0.171  &   0.111 & 0.406~  &  0.369 &  3.058 \\
PercCommercial    &   0.737 &  0.273  &   0.397 & 2.046~  &  0.355 &  3.073 \\
PercIndustrial    &   0.905 &  0.321  &   0.592 & 3.289*  &  0.323 &  3.215
\end{tabular}
}
\caption{\it LA Homeless data: Permutation Inference for $\RAVjhat$
  (10,000 permutations).  The value of $\RAVjhat$ for {\tt
    PercIndustrial} detects a statistically significant difference
  between $\SE_\lin$ and $\SE_\sand$ for this regressor.  See also
  Table~\ref{tab:Boston-RAV} in the Appendix for the Boston Housing
  data where significant differences are detected for 6 of 13
  regressors.  }
\label{tab:LA-RAV}
\end{table}

\subsection{An Approximate Permutation Distribution for the $\RAV$ Test Statistic}
\label{sec:RAV-test-permutation}

The asymptotic result of Proposition~\ref{sec:RAV-test-asymptotic}
provides qualitative insights, but it is not suitable for practical
application because the null distribution of $\RAVjhat$ can be very
non-normal for finite $N$, and this in ways that are not easily
overcome with simple tools such as nonlinear transformations.
Another approach to null distributions for finite $N$ is needed, and
it is available in the form of an approximate permutation test because
$H_0$ is just a null hypothesis of independence, here between
$\popres$ and $\Xjadj$.  The test is not exact, requiring
$N \!\gg\! p$, because population residuals $\popres_i$ must be
estimated with sample residuals $r_i$ and population adjusted
regressor values $\Xijadj$ with sample adjusted analogs $\Xijhadj$.
The permutation simulation is cheap: Once coordinate-wise squared
vectors $\Bresid^2$ and $\XBjhadj^2$ are formed, a draw from the
conditional null distribution of $\RAVjhat$ is obtained by randomly
permuting one of the vectors and forming the inner product with the
other, rescaled by a permutation-invariant factor
$N/(\|\Bresid\|^2 \|\XBjhadj\|^2)$.  A retention interval should be
formed directly from the $\alpha/2$ and $1\!-\!\alpha/2$ quantiles of
the permutation distribution to account for distributional
asymmetries.  The permutation distribution also yields an easy
diagnostic of non-normality (see Appendix~\ref{app:RAV-non-normality}
for examples).  Finally, by applying permutation simulations
simultaneously to $\RAV$ statistics of multiple regressors, one can
calibrate the retention intervals to control family-wise error.
---~See Table~\ref{tab:LA-RAV} (and \ref{tab:Boston-RAV} in the
Appendix) for examples of $\RAV$ tests.



\smallskip





\section{Issues with Model-Robust Standard Errors}
\label{sec:robustness}


Model-robustness is a highly desirable property, but as always there
is no free lunch.  Kauermann and Carroll (2001)\nocite{ref:KC-2001}
have shown that a cost of the sandwich estimator can be {\bf\em
  inefficiency when the assumed model is correct}.  Sandwich
estimators should be accurate only when the sample size is
sufficiently large.  

Another cost associated with the sandwich estimator is {\bf\em
  non-robustness in the sense of robust statistics} (Huber and
Ronchetti 2009\nocite{ref:HR-2009}, Hampel et
al. 1986\nocite{ref:HRRS-1986}), meaning strong sensitivity to
heavy-tailed distributions: The statistic $\SEhat_\sand^2[\betajhat]$
\eqref{eq:sandwich-AV-adj} is a ratio of fourth order quantities of
the data, whereas $\SEhat_\lin^2[\betajhat]$ \eqref{eq:SEhatlin} is
``only'' a ratio of second order quantities.\footnote{Note we are here
concerned with non-robustness of standard error estimates, not
parameter estimates.}
The two types of robustness are in conflict: Model-robust standard
error estimators are highly non-robust to heavy tails compared to
their model-trusting analogs.  This is a large issue which we can only
raise but not solve.  Here are some observations and suggestions:
\begin{itemize}
\item Classical robust regression may confer partial robustness to the
  sandwich standard error as it caps residuals with a bounded $\psi$
  function, thereby addressing robustness to heavy tails in the
  vertical ($y$) direction.  Anecdotal evidence suggests partial
  benefits.  In the LA Homeless data, for example, when comparing
  boostrap standard errors and standard errors reported by the {\em\bf
    R} (2008)\nocite{ref:R-2008}) software (function \texttt{lmrob} in
  package \texttt{robustbase}), we observed ratios
  $\frac{\SE_\boot}{\SE_\lmrob}$ of 1.470 and 0.957 for the
  coefficients of \texttt{PercVacant} and \texttt{PercIndustrial},
  respectively.  For linear OLS, the corresponding ratios
  $\frac{\SE_\boot}{\SE_\lin}$ in Table~\ref{tab:LA} were 1.513 and
  1.843, respectively.  Thus the roughly 50\% discrepancy for
  \texttt{PercVacant} persists, but the 80\% discrepancy for
  \texttt{PercIndustrial} is completely corrected.
\item Heavy-tail robustness in the horizontal ($\xvec$) direction can
  be achieved with bounded-influence regression (e.g., Krasker and
  Welsch 1982, and references therein\nocite{ref:KW-1982}) which
  downweights observations in high-leverage positions.
\item Robustness to horizontally heavy tails can also be addressed by
  transforming the regressor variables to bounded ranges (though this
  changes the meaning of the slopes).  Taking a cue from
  Proposition~\ref{app:extreme.RAV} in the appendix, one might search
  for transformations that obviate the need for a model-robust
  standard error in the first place.
\end{itemize}
To illustrate the last point, we transformed the regressors of the LA
Homeless data with their empirical cdfs to achieve approximately
uniform marginal distributions.  The transformed data are no longer
iid, but the point is to examine the effect of transforming the
regressors to a finite range.  As a result, shown in
Table~\ref{tab:LA-uniform} of
Appendix~\ref{app:LA-Data-CDF-transformed}, the discrepancies between
sandwich and usual standard errors have all but disappeared.  The same
drastic effect is not seen in the Boston Housing data
(Appendix~\ref{app:Boston}, Table~\ref{tab:Boston-uniform}), although
the discrepancies are greatly reduced, too.



%

\section{Summary and Outlook}
\label{sec:summary}

We explored for linear OLS the idea that statistical models imply
``simplification and idealization'' (Cox 1995)\nocite{ref:Cox-1995},
and hence should be treated as approximations rather than truths.  The
implications are many: (1)~Slope parameters need to be re-interpreted
as statistical functionals $\Bbeta(\PYX)$ arising from
best-approximating linear equations to essentially arbitrary
conditional mean functions $\mu(\Xvec)$; (2)~the presence of
nonlinearity $\nonlin(\Xvec)$ requires new interpretations of slope
parameters and their estimates; (3)~regressors are no longer ancillary
for the slope parameters; hence (4)~conditioning on the regressors is
not justified and regressors must be treated as random, arising from a
regressor distribution $\PX$; (5)~nonlinearity causes slope parameters
to depend not only on the conditional response distribution $\PYgX$
but on the regressor distribution $\PX$ as well; (6)~nonlinearity
causes randomness in the regressors $\Xvec$ to generate sampling
variation in slope estimates of order $N^{-1/2}$; (7)~sampling
variability due to $Y|\Xvec$ and due to $\Xvec$ are asymptotically
correctly captured by model-robust standard error estimates from the
$\xy$ bootstrap and sandwich plug-in, the latter being a limiting case
of the former; (8)~the factors that render the usual standard error of
a slope too liberal are strong nonlinearity and/or large noise
variance in the extremes of the adjusted regressor; (9)~validity of
the usual standard error varies from slope to slope but can be tested
with a slope-specific test; (10)~unresolved remains the problem that
model-robustness and classical heavy-tail robustness of standard error
estimates appear to be in conflict with each other.

A vexing item in this list is~(2): What is the meaning of a slope in
the presence of nonlinearity?  We gave an answer in terms of average
observed slopes, but this issue may remain controversial.  Yet, the
traditional interpretation of slopes should be even more controversial
because the notion of ``average difference in the response for a unit
difference in the regressor, ceteris paribus,'' tacitly assumes the
fitted linear equation to be correctly specified.  It remains correct
if ``in the response'' is replaced by ``in the best linear
approximation'', but this correction may leave some dissatified as
well.  Yet, misspecification is often a fact, as when simple models
are needed for substantive reasons or for communication with consumers
of statistical analysis.  It may then be prudent to use
interpretations and inferences that do not assume correct
specification.

Since White's seminal work, research into misspecification has
progressed far in addressing specific classes of misspecification:
dependencies, heteroskedasticities and nonlinearities.  A
generalization of White's sandwich estimator to time series dependence
in regression is the ``heteroskedasticity and auto-correlation
consistent'' (HAC) estimator of standard error by Newey and West
(1987)\nocite{ref:NW-1987}.  Structured second order misspecification
such as over/underdispersion have been addressed with
quasi-likelihood.  Intra-cluster dependencies in clustered (e.g.,
longitudinal) data have been addressed with generalized estimating
equations (GEE) where the sandwich estimator is in common use, as it
is in the generalized method of moments (GMM) literature.  Finally,
nonlinearities have been modeled with specific function classes or
estimated nonparametrically with, for example, additive models, spline
and kernel methods, and tree-based fitting.  In spite of these
advances, in finite data not all possibilities of misspecification can
be approached simultaneously, and there still arises a need for
model-robust inference.

There exist, finally, areas that frequently rely on model-trusting
theory:
\begin{itemize} \itemsep .3em
\item Bayes inference based on uninformative priors is asymptotically
  equivalent to model-trusting frequentist inference (Hartigan
  1983)\nocite{ref:Hartigan-1983}.  It should be reasonable to ask how
  much inferences from Bayesian models are adversely affected by
  misspecification.  After the early work by Berk (1966,
  1970)\nocite{ref:Berk-1966}\nocite{ref:Berk-1970} we find some more
  recent promising developments: Szpiro, Rice and Lumley
  (2010)\nocite{ref:SRL-2010} derive a sandwich estimator from
  Bayesian assumptions, and a lively discussion of misspecification
  from a Bayesian perspective involved Walker
  (2013)\nocite{ref:Walker-2013}, De Blasi
  (2013)\nocite{ref:DeBlasi-2013}, Hoff and Wakefield
  (2013)\nocite{ref:HW-2013} and O'Hagan
  (2013)\nocite{ref:OHagan-2013}, who provide further references.
\item High-dimensional inference is the subject of a large literature
  that often relies on the assumptions of linearity, homoskedasticity
  as well as normality of error distributions.  It may be uncertain
  whether procedures proposed in this area are model-robust.
  Recently, however, attention to the issue started to be paid by
  B\"uhlmann and van de Geer (2015)\nocite{ref:BvdG-2015}.  Relevant
  is also the incorporation of ideas from classical robust statistics
  by, for example, El Karoui et al. (2013)\nocite{ref:EBBY-2013},
  Donoho and Montanari (2014)\nocite{ref:DM-2014}, and
  Loh~(2015)\nocite{ref:Loh-2015}.
\end{itemize}
In summary, while interesting developments are in progress, there
remain open problems, especially in some of today's most lively
research areas.  Even in the non-Bayesian and low-dimensional domain
there remains the conflict between model-robustness and classical
robustness.  The implications of statistical models viewed as
approximations are not yet satisfactorily realized.



\newpage


\appendix


\section{LA Homeless data: SE discrepancies vanish after CDF transform of regressors}
\label{app:LA-Data-CDF-transformed}


\begin{table}[h]
\resizebox{\textwidth}{!}{\tt
\begin{tabular}{l|rrrrrrrrrr}
                  &$\betajhat$ & $\SE_\lin$ & $\SE_\boot$ & $\SE_\sand$ & $\frac{\SE_\boot}{\SE_\lin}$ & $\frac{\SE_\sand}{\SE_\lin}$ & $\frac{\SE_\sand}{\SE_\boot}$ & $t_\lin$ & $t_\boot$ & $t_\sand$ \\
\hline
(Intercept)    &   2.932 &  0.381 &  0.395  &  0.395  &   1.037  &    1.036  &    0.999&  7.697&  7.422&  7.427 \\
MedianIncome (\$K)&  -1.128 &  0.269 &  0.280  &  0.278  &   1.041  &    1.033  &    0.992& -4.195& -4.030& -4.061 \\
PercVacant     &   1.264 &  0.207 &  0.203  &  0.202  &   0.982  &    0.978  &    0.996&  6.111&  6.221&  6.247 \\
PercMinority   &  -0.467 &  0.230 &  0.246  &  0.246  &   1.070  &    1.069  &    0.999& -2.028& -1.896& -1.897 \\
PercResidential&  -0.314 &  0.220 &  0.228  &  0.230  &   1.040  &    1.049  &    1.008& -1.432& -1.377& -1.366 \\
PercCommercial &   0.201 &  0.212 &  0.220  &  0.220  &   1.040  &    1.042  &    1.002&  0.949&  0.913&  0.911 \\
PercIndustrial &   0.180 &  0.238 &  0.244  &  0.244  &   1.022  &    1.024  &    1.002&  0.754&  0.737&  0.736
\end{tabular}
}
\caption{\it LA Homeless Data: Comparison of Standard Errors after transforming
the regressors with their cdfs to approximately uniform distributions.  The
taming of the tails of the regressor distributions has resolved all discrepancy
issues for the usual model-trusting standard errors.}
\label{tab:LA-uniform}
\end{table}


\section{The Boston Housing data}
\label{app:Boston}

Table~\ref{tab:Boston} illustrates discrepancies between types of
standard errors with the Boston Housing data (Harrison and Rubinfeld
1978\nocite{ref:Boston-1978}) which will be well known to many
readers.  Again, we dispense with the question as to whether the
analysis is meaningful and focus on the comparison of standard errors.
Here, too, $\SE_\boot$ and $\SE_\sand$ are mostly in agreement as they
fall within less than 2\% of each other, an exception being
\texttt{CRIM} with a deviation of about 10\%.  By contrast,
$\SE_\boot$ and $\SE_\sand$ are larger than their linear models cousin
$\SE_\lin$ by a factor of about 2 for \texttt{RM} and \texttt{LSTAT},
and about 1.5 for the intercept and the dummy variable \texttt{CHAS}.
On the opposite side, $\SE_\boot$ and $\SE_\sand$ are less than 3/4 of
$\SE_\lin$ for \texttt{TAX}.  For several regressors there is no major
discrepancy among all three standard errors: \texttt{ZN},
\texttt{NOX}, \texttt{B}, and even for \texttt{CRIM}, $\SE_\lin$ falls
between the slightly discrepant values of $\SE_\boot$ and~$\SE_\sand$.

\begin{table}
\resizebox{\textwidth}{!}{\tt
\begin{tabular}{l|rrrrrrrrrr}
                  &$\betajhat$ & $\SE_\lin$ & $\SE_\boot$ & $\SE_\sand$ & $\frac{\SE_\boot}{\SE_\lin}$ & $\frac{\SE_\sand}{\SE_\lin}$ & $\frac{\SE_\sand}{\SE_\boot}$ & $t_\lin$ & $t_\boot$ & $t_\sand$ \\
\hline
\texttt{(Intercept)} &  36.459 & 5.103 & 8.038 & 8.145 & \bf 1.575 & \bf 1.596 & 1.013 &  7.144 &  4.536 &  4.477 \\
\texttt{CRIM}        &  -0.108 & 0.033 & 0.035 & 0.031 &     1.055 &     0.945 & 0.896 & -3.287 & -3.115 & -3.478 \\
\texttt{ZN}          &   0.046 & 0.014 & 0.014 & 0.014 &     1.005 &     1.011 & 1.006 &  3.382 &  3.364 &  3.345 \\
\texttt{INDUS}       &   0.021 & 0.061 & 0.051 & 0.051 & \bf 0.832 & \bf 0.823 & 0.990 &  0.334 &  0.402 &  0.406 \\
\texttt{CHAS}        &   2.687 & 0.862 & 1.307 & 1.310 & \bf 1.517 & \bf 1.521 & 1.003 &  3.118 &  2.056 &  2.051 \\
\texttt{NOX}         & -17.767 & 3.820 & 3.834 & 3.827 &     1.004 &     1.002 & 0.998 & -4.651 & -4.634 & -4.643 \\
\texttt{RM}          &   3.810 & 0.418 & 0.848 & 0.861 & \bf 2.030 & \bf 2.060 & 1.015 &  9.116 &  4.490 &  4.426 \\
\texttt{AGE}         &   0.001 & 0.013 & 0.016 & 0.017 &     1.238 &     1.263 & 1.020 &  0.052 &  0.042 &  0.042 \\
\texttt{DIS}         &  -1.476 & 0.199 & 0.214 & 0.217 &     1.075 &     1.086 & 1.010 & -7.398 & -6.882 & -6.812 \\
\texttt{RAD}         &   0.306 & 0.066 & 0.063 & 0.062 &     0.949 &     0.940 & 0.990 &  4.613 &  4.858 &  4.908 \\
\texttt{TAX}         &  -0.012 & 0.004 & 0.003 & 0.003 & \bf 0.736 & \bf 0.723 & 0.981 & -3.280 & -4.454 & -4.540 \\
\texttt{PTRATIO}     &  -0.953 & 0.131 & 0.118 & 0.118 &     0.899 &     0.904 & 1.005 & -7.283 & -8.104 & -8.060 \\
\texttt{B}           &   0.009 & 0.003 & 0.003 & 0.003 &     1.026 &     1.009 & 0.984 &  3.467 &  3.379 &  3.435 \\
\texttt{LSTAT}       &  -0.525 & 0.051 & 0.100 & 0.101 & \bf 1.980 & \bf 1.999 & 1.010 &-10.347 & -5.227 & -5.176
\end{tabular}
}
\caption{\it Boston Housing data: Comparison of Standard Errors.
  Sandwich and bootstrap $\SE$s are in general agreement, not so
  linear models $\SE$s.  While significances at conventional levels
  are unchanged in this case, the magnitude of the $t$ statistics can
  change drastically, for example, for {\tt RM} and {\tt LSTAT}, which
  are the strongest regressors.  }
\label{tab:Boston}
\end{table}

Table~\ref{tab:Boston-uniform} compares standard errors after the
regressors are transformed to approximately uniform distributions
using a rank or cdf transform.

\begin{table}
\resizebox{\textwidth}{!}{\tt
\begin{tabular}{l|rrrrrrrrrr}
                  &$\betajhat$ & $\SE_\lin$ & $\SE_\boot$ & $\SE_\sand$ & $\frac{\SE_\boot}{\SE_\lin}$ & $\frac{\SE_\sand}{\SE_\lin}$ & $\frac{\SE_\sand}{\SE_\boot}$ & $t_\lin$ & $t_\boot$ & $t_\sand$ \\
\hline
(Intercept) & 37.481 &  2.368 &  2.602 &   2.664  &   1.099  &    1.125  &    1.024&  15.828& 14.405& 14.069 \\
CRIM        &  4.179 &  1.746 &  1.539 &   1.533  &   0.882  &    0.878  &    0.996&   2.394&  2.715&  2.726 \\
ZN          &  0.826 &  1.418 &  1.359 &   1.353  &   0.959  &    0.954  &    0.995&   0.583&  0.608&  0.611 \\
INDUS       & -1.844 &  1.501 &  1.410 &   1.413  &   0.939  &    0.941  &    1.002&  -1.228& -1.308& -1.305 \\
CHAS        &  6.328 &  1.764 &  2.490 &   2.485  &\bf1.411  &\bf 1.409  &    0.998&   3.587&  2.542&  2.547 \\
NOX         & -6.209 &  1.986 &  2.035 &   2.037  &   1.025  &    1.026  &    1.001&  -3.127& -3.051& -3.048 \\
RM          &  4.848 &  1.044 &  1.354 &   1.380  &   1.297  &    1.322  &    1.019&   4.645&  3.581&  3.514 \\
AGE         &  2.925 &  1.454 &  1.897 &   1.904  &   1.305  &    1.310  &    1.004&   2.012&  1.542&  1.536 \\
DIS         & -9.047 &  1.754 &  1.933 &   1.945  &   1.102  &    1.109  &    1.006&  -5.159& -4.679& -4.652 \\
RAD         &  1.042 &  1.307 &  1.115 &   1.128  &   0.853  &    0.863  &    1.011&   0.797&  0.935&  0.924 \\
TAX         & -5.319 &  1.343 &  1.155 &   1.157  &   0.860  &    0.862  &    1.003&  -3.961& -4.607& -4.596 \\
PTRATIO     & -4.720 &  0.954 &  0.982 &   0.982  &   1.029  &    1.029  &    1.000&  -4.946& -4.806& -4.808 \\
B           & -1.103 &  0.822 &  0.798 &   0.800  &   0.970  &    0.972  &    1.002&  -1.342& -1.383& -1.380 \\
LSTAT       &-21.802 &  1.377 &  2.259 &   2.318  &\bf1.641  &\bf 1.683  &    1.026& -15.832& -9.649& -9.404
\end{tabular}
}
\caption{\it Boston Housing data: Comparison of Standard Errors;
  regressors are transformed with cdfs.  Forcing a bounded
  well-behaved distribution on the regressors greatly mitigates the
  discrepancies between $\SE$s.}
\label{tab:Boston-uniform}
\end{table}

Table~\ref{tab:Boston-RAV} illustrates the $\RAV$ test for the Boston
Housing data.  Values of $\RAVjhat$ that fall outside the middle 95\%
range of their permutation null distributions are marked with
asterisks.

\begin{table}
\resizebox{\textwidth}{!}{\tt
\begin{tabular}{l|rrrrrr}
                  &$\betajhat$ & $\SE_\lin$ & $\SE_\sand$ & $\RAVjhat$ & \textrm{2.5\% Perm.} & \textrm{97.5\% Perm.} \\
\hline
(Intercept)~~~~~~&  36.459 &  5.103  &   8.145 & 2.458* &     0.717 &  1.562   \\
CRIM             &  -0.108 &  0.033  &   0.031 & 0.776~ &     0.302 &  4.025   \\
ZN               &   0.046 &  0.014  &   0.014 & 1.006~ &     0.653 &  1.723   \\
INDUS            &   0.021 &  0.061  &   0.051 & 0.671~ &     0.641 &  2.007   \\
CHAS             &   2.687 &  0.862  &   1.310 & 2.255* &     0.499 &  1.929   \\
NOX              & -17.767 &  3.820  &   3.827 & 0.982~ &     0.694 &  1.579   \\
RM               &   3.810 &  0.418  &   0.861 & 4.087* &     0.631 &  1.805   \\
AGE              &   0.001 &  0.013  &   0.017 & 1.553* &     0.715 &  1.472   \\
DIS              &  -1.476 &  0.199  &   0.217 & 1.159~ &     0.702 &  1.520   \\
RAD              &   0.306 &  0.066  &   0.062 & 0.857~ &     0.685 &  1.983   \\
TAX              &  -0.012 &  0.004  &   0.003 & 0.512* &     0.579 &  1.998   \\
PTRATIO          &  -0.953 &  0.131  &   0.118 & 0.806~ &     0.725 &  1.395   \\  
B                &   0.009 &  0.003  &   0.003 & 0.995~ &     0.576 &  1.763   \\
LSTAT            &  -0.525 &  0.051  &   0.101 & 3.861* &     0.645 &  1.830 
\end{tabular}
}
\caption{\it Boston Housing data: Permutation Inference for $\RAVjhat$
  (10,000 permutations).  Of the 13 regressors, 6 have significant
  $\SE$ discrepancies, 5 of them indicating that linear models $\SE$s
  are too small, hence inferences are too optimistic.}
\label{tab:Boston-RAV}
\end{table}


\section{Ancillarity}
\label{app:ancillarity}

The facts as laid out in Section~\ref{sec:non-ancillarity} amount to
an argument against conditioning on regressors in regression.  The
justification for conditioning derives from an ancillarity argument
according to which the regressors, if random, form an ancillary
statistic for the linear model parameters $\Bbeta$ and $\sigma^2$,
hence conditioning on $\X$ produces valid frequentist inference for
these parameters (Cox and Hinkley 1974,
Example~2.27\nocite{ref:CH-1974}).  Indeed, with a suitably general
definition of ancillarity, it can be shown that in {\em any}
regression model the regressors form an ancillary.  To see this we
need an extended definition of ancillarity that includes nuisance
parameters.  The ingredients and conditions are as follows:
\begin{itemize}
\item[(1)] $\Btheta = (\Bpsi,\Blambda)\,$: the parameters, where $\Bpsi$
  is of interest and $\Blambda$ is nuisance;
\item[(2)] $\S = (\T,\A)\,$: a sufficient statistic with values $(\t,\a)$;
\item[(3)] $p(\t,\a;\,\Bpsi,\Blambda) = p(\t\,|\,\a;\,\Bpsi) \;
  p(\a;\,\Blambda)\,$: the condition that makes $\A$ an ancillary.
\end{itemize}
We say that the statistic $\A$ is ancillary for the parameter of
interest, $\Bpsi$, in the presence of the nuisance parameter,
$\Blambda$.  Condition (3) can be interpreted as saying that the
distribution of $\T$ is a mixture with mixing distribution
$p(\a|\Blambda)$.  More importantly, for a fixed but unknown value
$\Blambda$ and two values $\Bpsi_1$, $\Bpsi_0$, the likelihood ratio
\begin{equation*}
\frac{p(\t,\a;\,\Bpsi_1,\Blambda)}{p(\t,\a;\,\Bpsi_0,\Blambda)}
~=~
\frac{p(\t\,|\,\a;\,\Bpsi_1)}{p(\t\,|\,\a;\,\Bpsi_0)}
\end{equation*}
has the nuisance parameter $\Blambda$ eliminated, justifying the
conditionality principle according to which valid inference for
$\Bpsi$ can be obtained by conditioning on~$\A$.

When applied to regression, the principle implies that in {\em any}
regression model the regressors, when random, are ancillary and hence
can be conditioned on:
\begin{equation*}
  p(\y,\X;\,\Btheta) ~=~ p(\y\,|\,\X;\,\Btheta) \; p_\X(\X) ,
\end{equation*}
where $\X$ acts as the ancillary $\A$ and $p_\X$ as the mixing
distribution $p(\a\,|\Blambda)$ with a ``nonparametric'' nuisance
parameter that allows largely arbitrary distributions for the
regressors.  (The regressor distribution should grant identifiability
of $\Btheta$ in general, and non-collinearity in linear models in
particular.)  The literature does not seem to be rich in crisp
definitions of ancillarity, but see, for example, Cox and Hinkley
(1974, p.32-33)\nocite{ref:CH-1974}.  For the interesting history of
ancillarity see the articles by Stigler
(2001)\nocite{ref:Stigler-2001} and Aldrich
(2005)\nocite{ref:Aldrich-2005}.

As explained in Section~\ref{sec:non-ancillarity}, the problem with
the ancillarity argument is that it holds only when the regression
model is correct.  In practice, whether models are correct is never
known.


\section{Adjustment}
\label{app:adjustment}

\subsection{Adjustment in Populations}
\label{app:population-adjustment}

To define the population-adjusted regressor random variable $\Xjadj$,
collect all other regressors in the random $p$-vector
\begin{equation*}
  \Xvec_\noj = (1,\XXone,..., \XXjm, \XXjp,..., \XXp)\Tr,
\end{equation*}
and let
\begin{equation*}
  \Xjadj ~=~ X_j - \Xvec_{\noj}\Tr \Bbeta_{\noj \bull} , ~~~~~~
  \mbox{where}~~\Bbeta_{\noj \bull} = \E[\,\Xvec_\noj \Xvec_\noj\Tr]^{-1} \E[\,\Xvec_\noj X_j] .
\end{equation*}
The response $Y$ can be adjusted similarly, and we may denote it by
$Y_{\!\bull \, \noj}$ to indicate that $X_j$ is not among the
adjustors, which is implicit in the adjustment of~$X_j$.


\subsection{Adjustment in Samples}
\label{app:sample-adjustment}

Define the sample-adjusted regressor column $\XBjhadj$ by collecting
all regressor columns other than $\Xj$ in a $N \!\times\! p$ random
regressor matrix
\begin{equation*}
  \X_\noj = [\one, ..., \Xjm,\Xjp, ...,\Xp]
\end{equation*}
and let
\begin{equation*}
  \XBjhadj      ~=~ \Xj - \X_{\!\noj} \, \Bbetahat_{\noj \hbull}
  ~~~~~~ \hbox{where}~~
  \Bbetahat_{\noj \hbull}
  = (\X_\noj\Tr \X_\noj)^{-1} \X_\noj\Tr \Xj .
\end{equation*}
(Note the use of hat notation ``$\,\hbull\,$'' to distinguish it from
population-based adjustment ``$\bull$.'')  The response vector $\Y$
can be sample-adjusted similarly, and we may denote it by $\YBhadjj$
to indicate that $\Xj$ is not among the adjustors.


\section{Proofs}
\label{app:proofs}





\subsection{Precise Non-Ancillarity Statements and Proofs for Section~\ref{sec:non-ancillarity}}
\label{app:non-ancillarity}

\vspace{.1in}\noindent{\bf Lemma:~}{\em The functional
  $\BbetaP$ depends on $\P$ only through the conditional mean
  function and the regressor distribution; it does not depend on the
  conditional noise distribution.}

\smallskip

In the nonlinear case the clause $\exists \P_1, \P_2:
\Bbeta(\P_1) \neq \Bbeta(\P_2)$ is driven solely by differences in the
regressor distributions $\P_1(\d \xvec)$ and $\P_2(\d \xvec)$ because
$\P_1$ and $\P_2$ share the mean function $\mu_0(.)$ while their
conditional noise distributions are irrelevant by the above lemma.

The Lemma is more precisely stated as follows: For two data
distributions $\P_1(\d y, \d \xvec)$ and $\P_2(\d y, \d \xvec)$ the
following holds:
\begin{equation*}
  \P_1(\d \xvec) = \P_2(\d \xvec),~~~~
  \mu_1(\Xvec) \overset{\P_{1,2}}{=} \mu_2(\Xvec) 
  ~~~~ \Longrightarrow ~~~~
  \Bbeta(\P_1) = \Bbeta(\P_2).
\end{equation*}

\vspace{.1in}\noindent{\bf Proposition:~}{\em The OLS functional $\BbetaP$
  does {\bf not} depend on the regressor distribution if and only if
  $\mu(\Xvec)$ is linear.  More precisely, for a fixed measurable
  function $\mu_0(\xvec)$ consider the class of data distributions
  $\P$ for which $\mu_0(.)$ is a version of their conditional mean
  function: $\E[Y|\Xvec] = \mu(\Xvec) \overset{\as}{=} \mu_o(\Xvec)$.
  In this class the following holds:}
\begin{equation*}
  \begin{array}{lll}
    \mu_0(.) ~{is~nonlinear} &
    ~~~~ \Longrightarrow ~~~~  &
    \exists \P_1, \P_2: ~~ \Bbeta(\P_1) \neq \Bbeta(\P_2) , \\
    \mu_0(.) ~{is~linear~~~~~} &
    ~~~~ \Longrightarrow ~~~~ &
    \forall \P_1, \P_2: ~~ \Bbeta(\P_1)  =  \Bbeta(\P_2) .
  \end{array}
\end{equation*}

For the proposition we show the following: For a fixed measurable
function $\mu_0(\xvec)$ consider the class of data distributions $\P$
for which $\mu_0(.)$ is a version of their conditional mean function:
$\E[Y|\Xvec] = \mu(\Xvec) \overset{\as}{=} \mu_o(\Xvec)$.  In this
class the following holds:
\begin{equation*}
  \begin{array}{lll}
    \mu_0(.) ~{is~nonlinear} &
    ~~~~ \Longrightarrow ~~~~  &
    \exists \P_1, \P_2: ~~ \Bbeta(\P_1) \neq \Bbeta(\P_2) , \\
    \mu_0(.) ~{is~linear~~~~~} &
    ~~~~ \Longrightarrow ~~~~ &
    \forall \P_1, \P_2: ~~ \Bbeta(\P_1)  =  \Bbeta(\P_2) .
  \end{array}
\end{equation*}

The linear case is trivial: if $\mu_0(\Xvec)$ is linear, that is,
$\mu_0(\xvec) = \Bbeta\Tr \xvec$ for some $\Bbeta$, then $\BbetaP =
\Bbeta$ irrespective of $\P(\d \xvec)$.
The nonlinear case is proved as follows: For
any set of points $\xvec_1,... \xvec_{p\!+\!1} \in \Reals^{p\!+\!1}$
in general position and with 1 in the first coordinate, there exists a
unique linear function $\Bbeta\Tr \xvec$ through the values of
$\mu_0(\xvec_i)$.  Define $\P(\d \xvec)$ by putting mass $1/(p\!+\!1)$
on each point; define the conditional distribution $\P(\d y \,|\,
\xvec_i)$ as a point mass at $y = \mu_o(\xvec_i)$; this defines $\P$
such that $\BbetaP = \Bbeta$.  Now, if $\mu_0()$ is nonlinear,
there exist two such sets of points with differing linear functions
$\Bbeta_1\Tr \xvec$ and $\Bbeta_2\Tr \xvec$ to match the values of
$\mu_0()$ on these two sets; by following the preceding construction
we obtain $\P_1$ and $\P_2$ such that $\Bbeta(\P_1) = \Bbeta_1 \neq
\Bbeta_2 = \Bbeta(\P_2)$.





\subsection{$\RAV$ Decomposition}
\label{app:RAV-decomposition}

\noindent{\bf Lemma \ref{app:RAV-decomposition}:}{\it~~$\RAV\!$ Decomposition.}
\begin{equation*} 
  \begin{array}{l}
    \DS \RAV[\betajhat,\rmse^2]
    ~=~
    w_{\sigma} \, \RAV[\betajhat,\sigma^2]
    ~+~
    w_{\nonlin} \, \RAV[\betajhat,\nonlin^2] ,
    \\ \\
    \textit{where}~~~~
    \DS
    w_{\sigma} ~\defn~ \frac{E[\sig^2(\Xvec)]}{\E[\rmse^2(\Xvec)]} \,,
    ~~~
    w_{\nonlin} ~\defn~ \frac{\E[\nonlin^2(\Xvec)]}{\E[\rmse^2(\Xvec)]} \,,
    ~~~
    w_{\sigma} + w_{\nonlin} = 1.
  \end{array}
\end{equation*}

\vspace{.05in}

\subsection{Proof of the $\RAV$-Range Proposition in Section~\ref{sec:range-of-RAV}}
\label{app:extreme.RAV}



\noindent{\bf Proposition~\ref{app:extreme.RAV}:} {\em
If $\E[\Xjadjsq] < \infty$, then
\begin{equation*}
  \sup_{\rmse_j^2} \RAV[\betajhat,\rmse_j^2] ~=~ \frac{\Pmax \Xjadjsq}{\E[\Xjadjsq]} ,
  ~~~~~~
  \inf_{\rmse_j^2} \RAV[\betajhat,\rmse_j^2] ~=~ \frac{\Pmin \Xjadjsq}{\E[\Xjadjsq]} .
\end{equation*}
}

\bigskip

Here are some corollaries that follow from the proposition:
\begin{itemize} \itemsep 0.5em
\item If, for example, $\Xjadj \sim U[-1,+1]$ is uniformly
  distributed, then $\E[\Xjadjsq] = 1/3$.  Hence the upper bound on
  the $\RAV$ is 3 and, asymptotically, the usual standard error will
  never be too short by more than a factor $\sqrt{3} \approx 1.732$.
\item However, when $\E[\Xjadjsq]$ is very small compared to
  $\Pmax \Xjadjsq$, that is, when $\Xjadj$ is highly concentrated around
  its mean 0, then this approximates the case of an unbounded support
  and the worst-case $\RAV$ can be very large.
\item If, on the other hand, $\E[\Xjadjsq]$ is very close to
  $\,\Pmax \Xjadjsq = c^2$, then $\Xjadj$ approximates a balanced
  two-point distribution at $\pm c$, and the sandwich and usual
  standard errors necessarily agree in the limit.
\end{itemize}
The result for the last case, a two-point balanced distribution, is
intuitive because here it is impossible to detect nonlinearity.
Heteroskedasticity, however, is still possible (different noise
variances at $\pm c$), but this does not matter because the dependence
of $\RAV$ is on $\Xjadjsq$, not $\Xjadj$, and $\Xjadjsq$ has a
one-point distribution at~$c^2$.  The $\RAV$ can only respond to
heteroskedasticities that vary in~$\Xjadjsq$.

The $\RAV$ is a functional of $\Xjadjsq$ and $f_j^2(\Xjadjsq)$,
suggesting simplified notation: $\Xsq$ for $\Xjadjsq$, $\fsq(\Xsq)$
for~$f_j^2(\Xjadjsq)$, and $\RAV[\fsq]$ for $\RAV[\betajhat,f_j^2]$.
Proposition~\ref{app:extreme.RAV} is proved by the first lemma as
applied to $\sig_j^2(\Xjadjsq)$, and by the second lemma as applied to
$\nonlin_j^2(\Xjadjsq)$.  The difference between the two cases is that
nonlinearities $\nonlin_j(\Xjadjsq)$ is necessarily centered whereas
for~$\sig_j^2(\Xjadjsq)$ there exists no such requirement; the
construction below requires in the centered case that $\Pmin$ and
$\Pmax$ of $\Xjadjsq$ do not carry positive probability mass.  This is
a largely technical condition because even for discrete regressors
$\XX_j$ the adjusted squared version $\Xjadjsq$ will have a continuous
distribution if there exists just one other regressor that is
continuous and non-orthogonal (partly collinear) to~$\XXj$.

\medskip

\noindent{\bf Lemma  \ref{app:extreme.RAV}.1:}
{\it Assume $\E[\Xsq] < \infty$.
  \begin{itemize}
    \item[(a)] Define a one-parameter family $\fsqt$:
      \begin{equation*} \label{eq:RAV-nonlinear}
        \fsqt(\Xsq) ~~\defn~~ \frac{1_{[|X| \ge t]}}{\p(t)}
        \,, ~~~~~~{where}~~~~\p(t) ~\defn~ \P[|X| \ge t]
      \end{equation*}
      for $\p(t) > 0$.  Then the following holds:
      \begin{equation*}
        \sup_t \RAV[\fsqt] ~=~ \frac{\Pmax \Xsq}{\E[\Xsq]} .
      \end{equation*}

    \item[(b)] Define a one-parameter family $\gsqt$:
      \begin{equation*}
        \gsqt(\Xsq) ~~\defn~~ \frac{1_{[|X| \le t]}}{\pbar(t)}
        \,, ~~~~~~{where}~~~~\pbar(t) ~\defn~ \P[|X| \le t]\,.
      \end{equation*}
      Then the following holds:
      \begin{equation*}
        \inf_t \RAV[\gsqt] ~=~ \frac{\Pmin \Xsq}{\E[\Xsq]} .
      \end{equation*}
    \end{itemize}
  }

\smallskip

\noindent{\bf Proof of part (a):} Preliminary observations:
\begin{itemize} \itemsep .4em
  \item $\E[\fsqt(\Xsq)] = 1$.
  \item $\E[\fsqt(\Xsq) \Xsq] \le \Pmax \Xsq$.
  \item $\Pmax \Xsq = \sup_{\p(t) > 0} t^2$.
\end{itemize}
For $\p(t) > 0$ we have
  \begin{equation*}
    \E\left[ \fsqt(X) X^2 \right]
    ~=~ \frac{1}{\p(t)} \, \E\left[ 1_{[|X| \ge t]} \, X^2 \right]
    ~\ge~ \frac{1}{\p(t)} \, \p(t) \, t^2
    ~=~ t^2 ,
  \end{equation*}
hence~ $\sup_t \E\left[ \fsqt(X) X^2 \right] = \Pmax \Xsq$. ~~~$\square$

\bigskip

\noindent{\bf Proof of part (b):} Preliminary observations:
\begin{itemize} \itemsep .4em
  \item $\E[\gsqt(\Xsq)] = 1$.
  \item $\E[\gsqt(\Xsq) \Xsq] \ge \Pmin \Xsq$.
  \item $\Pmin \Xsq = \inf_{\pbar(t) > 0} t^2$.
\end{itemize}
For $\pbar(t) > 0$ we have:
  \begin{equation*}
    \E\left[ \gsqt(X) X^2 \right]
    ~=~ \frac{1}{\pbar(t)} \, \E\left[ 1_{[|X| \le t]} \, X^2 \right]
    ~\le~ \frac{1}{\pbar(t)} \, \pbar(t) \, t^2
    ~=~ t^2 ,
  \end{equation*}
hence $\inf_t \E\left[ \gsqt(X) X^2 \right] = \Pmin \Xsq$. ~~~$\square$

\vspace{2.5em}

\noindent{\bf Lemma \ref{app:extreme.RAV}.2:} {\it
  \begin{itemize}
  \item[(a)] Define a one-parameter family

    \begin{equation*}
      \ft(\Xsq) ~=~ \frac{1_{[|X| \ge t]} - \p(t)}{\sqrt{\p(t)(1-\p(t))}}
      \,, ~~~~~~{where}~~~~\p(t) = \P[|X| \ge t]\,,
    \end{equation*}
    for $\p(t) \!>\! 0$ and $1 \!-\! \p(t) \!>\! 0$.  If $p(t)$ is continuous at $t \!=\! \Pmax |X|$,
    that is, $\P[|X| = \Pmax |X|] = 0$, then
    \begin{equation*}
      \sup_t \, \RAV[\fsqt] ~=~ \frac{\Pmax \Xsq}{\E[\Xsq]} .
    \end{equation*}

  \item[(b)] Define a one-parameter family
    \begin{equation*}
      \gt(\Xsq) ~=~ \frac{1_{[|X| \le t]} - \pbar(t)}{\sqrt{\pbar(t)(1-\pbar(t))}}
      \,, ~~~~~~{where}~~~~\pbar(t) = \P[|X| \le t]\,,
    \end{equation*}
    for $\pbar(t) \!>\! 0$ and $1 \!-\! \pbar(t) \!>\! 0$.  If $\pbar(t)$ is continuous at $t \!=\! \Pmin |X|$,
    that is, $\P[|X| = \Pmin |X|] = 0$, then
    \begin{equation*}
      \inf_t \, \RAV[\gsqt] ~=~ \frac{\Pmin \Xsq}{\E[\Xsq]} .
    \end{equation*}

  \end{itemize}
}

\smallskip

\noindent{\bf Proof of part (a):} Preliminary observations:
\begin{itemize} \itemsep .4em
  \item $\E[\fsqt(\Xsq)] = 1$.
  \item $\E[\fsqt(\Xsq) \Xsq] \le \Pmax \Xsq$.
  \item $\Pmax \Xsq = \sup_{\,0< \p(t) <1} t^2$.
\end{itemize}
For $\p(t) \!>\! 0$ we have:
  \begin{align*}
    \E\left[ \fsqt(X) X^2 \right]
    &~=~ \frac{1}{\p(t)(1-\p(t))} \, \E\left[ \left( 1_{[|X| \ge t]} - \p(t) \right)^2 X^2 \right]
    \\
    &~=~ \frac{1}{\p(t)(1-\p(t))} \, \left( \E\left[ 1_{[|X| \ge t]} X^2 \right] (1 - 2 \, \p(t)) + \p(t)^2  \E[X^2] \right)
    \\
    &~\ge~ \frac{1}{\p(t)(1-\p(t))} \, \left( \p(t) \, t^2 \, (1 - 2 \, \p(t)) \, + \p(t)^2  \E[X^2]\right)
                 ~~~~~~{\rm for}~~~~\p(t)\le \frac{1}{2}
    \\
    &~=~ \frac{1}{1-\p(t)} \, \left( t^2 \, (1 - 2 \, \p(t))  + \p(t)  \E[X^2]\, \right)
    \\
    & \longrightarrow ~ \Pmax \Xsq
  \end{align*}
as~ $t \uparrow \Pmax |X|$ ~~and hence~ $\p(t) \downarrow 0$. ~~~$\square$

\bigskip

\noindent{\bf Proof of part (b):} Preliminary observations:
\begin{itemize} \itemsep .4em
  \item $\E[\gsqt(\Xsq)] = 1$.
  \item $\E[\gsqt(\Xsq) \Xsq] \ge \Pmin \Xsq$.
  \item $\Pmin \Xsq = \inf_{\,0< \pbar(t) <1} t^2$.
\end{itemize}
  \begin{align*}
    \E\left[ \gsqt(X)^2 X^2 \right]
    &~=~ \frac{1}{\pbar(t)(1-\pbar(t))} \, \E\left[ \left( 1_{[|X| \le t]} - \pbar(t) \right)^2 X^2 \right]
    \\
    &~=~ \frac{1}{\pbar(t)(1-\pbar(t))} \, \left( \E\left[ 1_{[|X| \le t]} X^2 (1 - 2 \, \pbar(t)) \right] + \pbar(t)^2 \E[X^2] \right)
    \\
    &~\le~ \frac{1}{\pbar(t)(1-\pbar(t))} \, \left( \pbar(t) \, t^2 \, (1 - 2 \, \pbar(t)) + \pbar(t)^2 \E[X^2] \right)
                 ~~~~~~{\rm for}~~~~\pbar(t)\le \frac{1}{2}
    \\
    &~=~ \frac{1}{1-\pbar(t)} \, \left( t^2 \, (1 - 2 \, \pbar(t)) + \pbar(t) \E[X^2] \, \right)
    \\
    & \longrightarrow ~ \Pmin \Xsq
  \end{align*}
as~ $t \downarrow \Pmin |X|$ ~~and hence~ $\pbar(t) \downarrow 0$. ~~~$\square$

\bigskip


\subsection{Details for Figure~\ref{fig:RAV-f}}
\label{app:examples-for-sig-and-eta}

We write $X$ instead of $\Xjadj$ and assume it has a standard
normal distribution, $X \sim N(0,1)$, whose density will be denoted by
$\phi(x)$.  In Figure~\ref{fig:RAV-f} the base
function is, up to scale, as follows:
\begin{equation*}
  f(x) ~=~ \exp\left( -\frac{t}{2} \, \frac{\x^2}{2} \right),~~~~~~ t > -1 .
\end{equation*}
These functions are normal densities up to normalization for $t>0$,
constant 1 for $t=0$, and convex for $t < 0$.  Conveniently, $f(x)
\phi(x)$ and $f^2(x)\phi(x)$ are both normal densities (up to
normalization) for $t > -1$:
\begin{equation*}
  \begin{array}{lcll}
    f(x) \, \phi(x)   &=& \siga \, \phi_\siga(x) , ~~~~~~~ &\siga = (1+t/2)^{-1/2} , \\
    f^2(x) \, \phi(x) &=& \sigb \, \phi_\sigb(x) , ~~~~~~~ &\sigb = (1+t)^{-1/2} ,
  \end{array}
\end{equation*}
where we write $\phi_s(x) = \phi(x/s)/s$ for scaled normal
densities.  Accordingly we obtain the following moments:
\begin{equation*}
  \begin{array}{lclclcl}
    \E[f(X)]          &=& \siga \, \E[\,1\, | N(0,\siga^2) ] &=& \siga   &=& (1+t/2)^{-1/2} , \\
    \E[f(X) \, X^2]   &=& \siga \, \E[X^2   | N(0,\siga^2) ]  &=& \siga^3 &=& (1+t/2)^{-3/2} , \\
    \E[f^2(X)]        &=& \sigb \, \E[\,1\, | N(0,\sigb^2) ]  &=& \sigb   &=& (1+t)^{-1/2}   , \\
    \E[f^2(X) \, X^2] &=& \sigb \, \E[X^2    | N(0,\sigb^2) ]  &=& \sigb^3 &=&(1+t)^{-3/2}    ,
  \end{array}
\end{equation*}
and hence
\begin{equation*}
    \RAV[\betahat,f^2] ~=~ \frac{ \E[f^2(X) \, X^2] } {\E[f^2(X)] \, \E[X^2] }
                 ~=~ \sigb^2 ~=~ (1+t)^{-1}
\end{equation*}
Figure \ref{fig:RAV-f} shows the functions as follows:
  $f(x)^2/\E[f^2(X)] = f(x)^2/\sigb$.


\subsection{Proof of Asymptotic Normality of $\RAVjhat$, Section~\ref{sec:RAV-test-asymptotic}}
\label{app:RAV-test}

We will need notation for each observation's population-adjusted
regressors:
$\X_{\!j \bull} = (X_{1,j\bull},...,X_{N,j\bull})\Tr = \Xj - \X_{\!\noj}
\Bbeta_{\noj \bull}$.
The following distinction is elementary but important: The component
variables of $\XBjadj=(\Xijadj)_{i=1...N}$ are iid~as they are
population-adjusted, whereas the component variables of
$\XBjhadj=(\Xijhadj)_{i=1...N}$ are dependent as they are
sample-adjusted.  As $N \rightarrow \infty$ for fixed $p$, this
dependency disappears asymptotically, and we have for the empirical
distribution of the values $\{\Xijhadj\}_{i=1...N}$ the obvious
convergence in distribution:
\begin{equation*}
  \{\Xijhadj\}_{i=1...N} ~ ~~\stackrel{\Dist}{\longrightarrow}~~ ~ \Xjadj
  ~ \overset{\Dist}{=} ~ \Xijadj
  ~~~~~~~~(N \rightarrow \infty) .
\end{equation*}

We recall \eqref{eq:RAV-est} for reference in the following form:
\begin{equation} \label{eq:RAV-again}
  \RAVjhat ~=~ \frac{ \frac{1}{N} \langle (\Y -\X \Bbetahat)^2, \XBjhadj^2 \rangle }
                    {  \frac{1}{N} \| \Y -\X \Bbetahat \|^2 \;  \frac{1}{N} \| \XBjhadj \|^2 } \,.
\end{equation}
For the denominators it is easy to show that
\begin{equation} \label{eq:RAVhat-denom}
  \begin{array}{rcl}
    \frac{1}{N} \| \Y -\X \Bbetahat \|^2
    &\overset{\P}{\longrightarrow}&
    \E[\,\popres^2\,] ,
    \\[2pt]
    \frac{1}{N} \| \XBjhadj \|^2
    &\overset{\P}{\longrightarrow}&
    \E[\,\Xjadjsq\,] .
  \end{array}
\end{equation}
For the numerator a CLT holds based on
\begin{eqnarray}
    {\TS \frac{1}{N^{1/2}}} \langle (\Y -\X \Bbetahat)^2, \XBjhadj^2 \rangle
    & = & \label{eq:RAVhat-num}
    {\TS \frac{1}{N^{1/2}}} \langle (\Y -\X \Bbeta)^2, \XBjadj^2 \rangle + O_P(N^{-1/2}) .
\end{eqnarray}
For a proof outline see {\bf Details} below.  It is therefore
sufficient to show asymptotic normality of $\langle \Bpopres^2,
\XBjadj^2 \rangle$.  Here are first and second moments:
\begin{equation*}
  \begin{array}{rclcl}
    \E[\frac{1}{N} \langle \Bpopres^2, \XBjadj^2 \rangle]
    & = &
    \E[\popres^2 \, \Xjadjsq]
    & = &
    \E[\popres^2] \, \E[\Xjadjsq] ,
    \\[4pt]
    \V[\frac{1}{N^{1/2}} \langle \Bpopres^2, \XBjadj^2 \rangle]
    & = &
    \E[\popres^4 \, \Xjadj^4] -  \E[\popres^2 \, \Xjadjsq]^2
    & = &
    \E[\popres^4] \, \E[\Xjadj^4] - \E[\popres^2]^2 \, \E[\Xjadjsq]^2 .
  \end{array}
\end{equation*}
The second equality on each line holds under the null hypothesis of
independent $\popres$ and $\Xvec$.  For the variance one observes that
we assume that $\{(Y_i,\Xveci)\}_{i=1...N}$ to be iid~sampled
pairs, hence $\{(\popres_i^2,\Xijadj^2)\}_{i=1...N}$ are $N$
iid~sampled pairs as well.  Using the denominator terms
\eqref{eq:RAVhat-denom} and Slutsky's theorem, we arrive at the first
version of the CLT for~$\RAVjhat$:
\begin{equation*}
  \begin{array}{rcl}
    N^{1/2} \, ( \RAVjhat - 1 )
    & \overset{\Dist}{\longrightarrow} &
    \Norm\left( 0,\, {\DS \frac{\E[\,\popres^4]}
                               {\E[\,\popres^2]^2} \,
                          \frac{\E[\Xjadj^4]}
                               {\E[\Xjadjsq]^2 }
                    \,-\, 1 }
         \right)
  \end{array}
\end{equation*}
With the additional null assumption of normal noise we have
$\E[\,\popres^4] = 3 \E[\,\popres^2]^2$, and hence the second version
of the CLT for~$\RAVjhat$:
\begin{equation*}
  \begin{array}{rcl}
    N^{1/2} \, ( \RAVjhat - 1 )
    & \overset{\Dist}{\longrightarrow} &
    \Norm\left( 0,\, 3 \, {\DS \frac{\E[\Xjadj^4]}
                               {\E[\Xjadjsq]^2 }
                    - 1 }
         \right) .
  \end{array}
\end{equation*}

\bigskip

\noindent {\bf Details for the numerator} \eqref{eq:RAVhat-num}, using
notation of Sections~\ref{app:population-adjustment}
and~\ref{app:sample-adjustment}, in particular $\X_{\!j \bull} = \Xj -
\X_{\noj} \Bbeta_{\noj \bull}$ and $\X_{\!j \hbull} = \Xj - \X_{\!\noj}
\Bbetahat_{\noj \hbull}$:
\begin{equation} \label{eq:numerator}
  \begin{array}{rcl}
    \langle (\Y -\X \Bbetahat)^2, \XBjhadj^2 \rangle
    & = &
    \langle\,
      ((\Y -\X \Bbeta) - \X (\Bbetahat - \Bbeta))^2
      ,\,
      (\XBjadj - \X_\noj (\Bbetahat_{\noj \hbull} - \Bbeta_{\noj \bull}) )^2
    \,\rangle
    \\[4pt]
    & = &
    \langle\,
      \Bpopres^2
      + (\X (\Bbetahat - \Bbeta))^2
      - 2\, \Bpopres \, (\X (\Bbetahat - \Bbeta))
      ,\,
      \\
      & & ~~
      \XBjadj^2
      + (\X_\noj (\Bbetahat_{\noj \hbull} - \Bbetahat_{\noj \bull}) )^2
      - 2\, \XBjadj ( \X_\noj (\Bbetahat_{\noj \hbull} - \Bbeta_{\noj \bull}) )
    \,\rangle
    \\[4pt]
    & = &
    \langle\,
      \Bpopres^2, \XBjadj^2
    \,\rangle
    + ...
  \end{array}
\end{equation}
Among the 8 terms in ``...'', each contains at least one subterm of
the form $\Bbetahat - \Bbeta$ or $\Bbetahat_{\noj \hbull} -
\Bbeta_{\noj \bull}$, each being of order $O_P(N^{-1/2})$.  We first
treat the terms with just one of these subterms to first power, of
which there are only two, normalized by $N^{1/2}$:
\begin{equation*}
  \begin{array}{rcl}
    \frac{1}{N^{1/2}} \,
    \langle\,
      - 2\, \Bpopres \, (\X (\Bbetahat - \Bbeta))
      ,\,
      \XBjadj^2
    \,\rangle
    &=&
    -2 \,
    \sum_{k=0...p} \,
      \left(
        \frac{1}{N^{1/2}}
        \sum_{i=1...N}
          \popres_i X_{i,k} X_{i,j\bull}^2
      \right)
      (\betajhat - \betaj)
    \\[4pt]
    &=& \sum_{k=0...p} \, O_P(1) \, O_P(N^{-1/2})  ~=~ O_P(N^{-1/2}),
    \\[8pt]
    \frac{1}{N^{1/2}} \,
    \langle\,
      \Bpopres^2
      ,\,
      - 2\, \XBjadj ( \X_\noj (\Bbetahat_{\noj \hbull} - \Bbeta_{\noj \bull}) )
    \,\rangle
    &=&
    -2 \,
      \sum_{k (\neq j)} \,
        \left(
          \frac{1}{N^{1/2}}
          \sum_{i=1...N}
             \popres_i^2 X_{i,j\bull} X_{i,k}
        \right)
        (\betahat_{\noj \hbull,k} - \beta_{\noj \bull,k})
    \\[4pt]
    &=&
    \sum_{k (\neq j)} \, O_P(1) \, O_P(N^{-1/2}) ~=~ O_P(N^{-1/2}).
  \end{array}
\end{equation*}
The terms in the big parens are $O_P(1)$ because they are
asymptotically normal.  This is so because they are centered under the
null hypothesis that $\popres_i$ is independent of the regressors
$\Xveci$: In the first term we have
\begin{equation*}
  \E[\popres_i X_{i,k} X_{i,j\bull}^2] = \E[\popres_i] \, \E[X_{i,k} X_{i,j\bull}^2] = 0
\end{equation*}
due to $\E[\popres_i] = 0$.  In the second term we have
\begin{equation*}
  \E[\popres_i^2 X_{i,j\bull} X_{i,k}] = \E[\popres_i^2] \, \E[X_{i,j\bull} X_{i,k}] = 0
\end{equation*}
due to $\E[X_{i,j\bull} X_{i,k}] = 0$ as~$k \neq j$.

We proceed to the 6 terms in \eqref{eq:numerator} that contain at
least two $\beta$-subterms or one $\beta$-subterm squared.  For
brevity we treat one term in detail and assume that the reader will be
convinced that the other 5 terms can be dealt with similarly.  Here is
one such term, again scaled for CLT purposes:
\[
  \begin{array}{rcl}
  \frac{1}{N^{1/2}}
  \langle\,
    (\X(\Bbetahat - \Bbeta))^2, \XBjadj^2
  \,\rangle
  &=&
  \sum_{k,l=0...p}
    \left(
      \frac{1}{N}
      \sum_{i=1...N}
        X_{i,k} X_{i,l} X_{i,j\bull}^2
    \right)
    N^{1/2} (\betahat_k - \beta_k)(\betahat_l - \beta_l)
  \\
  &=&
  \sum_{k,l=0...p} \const \cdot O_P(1) \, O_P(N^{-1/2}) ~=~ O_P(N^{-1/2}).
  \end{array}
\]
The term in the parens converges in probability to $\E[X_{i,k} X_{i,l}
X_{i,j\bull}^2 ]$, accounting for ``const''; the term $N^{1/2}
(\betahat_k - \beta_k)$ is asymptotically normal and hence $O_P(1)$;
and the term $(\betahat_l - \beta_l)$ is $O_P(N^{-1/2})$ due to its
CLT.

\noindent{\bf Details for the denominator
  terms}~\eqref{eq:RAVhat-denom}: It is sufficient to consider the
first denominator term.  Let $\H = \X (\X\Tr \X)^{-1} \X\Tr$ be the
hat or projection matrix for~$\X$.
\begin{align*}
  {\TS \frac{1}{N}}\, \|\Y - \X \Bbetahat\|^2
    &~=~ {\TS \frac{1}{N}}\, \Y\Tr (\I - \H) \Y
    \\
    &~=~ {\TS \frac{1}{N}}\, \left( \|\Y\|^2 - \Y\Tr \H \Y \right)
    \\
    &~=~ {\TS \frac{1}{N}}\, \|\Y\|^2  ~-~
        \left( {\TS \frac{1}{N}} \sum Y_{i} \Xvec_{i}\Tr \right)
        \left( {\TS \frac{1}{N}} \sum \Xvec_{i} \Xvec_{i}\Tr \right)^{-1}
        \left( {\TS \frac{1}{N}} \sum \Xvec_{i}  Y_{i} \right)
    \\
    &\stackrel{\P}{\longrightarrow}~
    \E[Y^2] ~-~ \E[Y \Xvec] \, \E[\Xvec \Xvec\Tr]^{-1} \E[\Xvec Y]
    \\
    &~=~
    \E[Y^2] - \E[Y \Xvec\Tr \Bbeta]
    \\
    &~=~
    \E[ (Y - \Xvec\Tr \Bbeta)^2 ] ~~~~~~~~{\rm due~to}~~\E[(Y-\Xvec\Tr \Bbeta) \Xvec] = \zero
    \\
    &~=~
    \E[\,\popres^2] .
\end{align*}
The calculations are the same for the second denominator term,
substituting $\Xj$ for $\Y$, $\X_{\!\noj}$ for $\X$, $X_{\!j \bull}$
for $\popres$, and $\Bbeta_{\noj \bull}$ for~$\Bbeta$.


\newpage

\section{Non-Normality of Conditional Null Distributions of $\RAVjhat$}
\label{app:RAV-non-normality}

\begin{figure}[h]
  \centering
  \includegraphics[width=5.in]{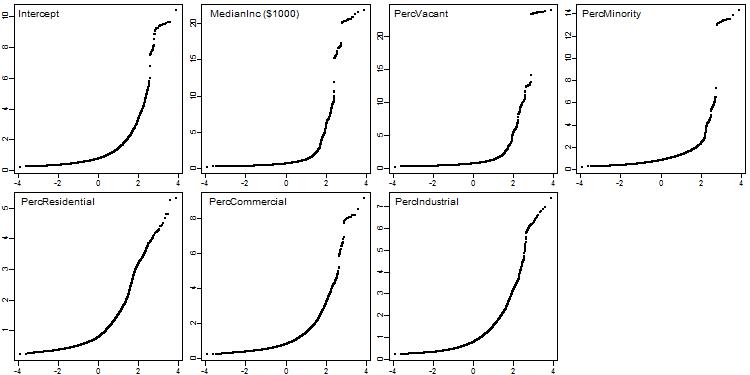}
  \caption{Permutation distributions of $\RAVjhat$ for the LA Homeless Data}
\end{figure}

\vspace{.3in}

\begin{figure}[h]
  \centering
  \includegraphics[width=5.in]{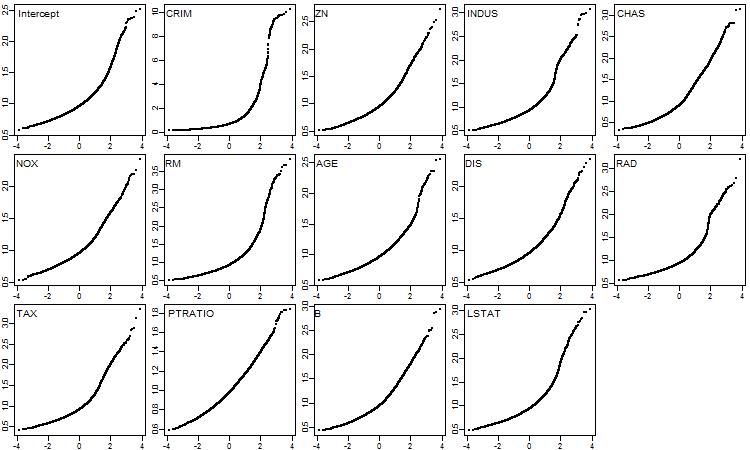}
  \caption{Permutation distributions of $\RAVjhat$ for the Boston Housing Data}
\end{figure}


\end{document}